\documentclass[10pt, reprint, twocolumn, amssymb, amsmath, tightenlines, footinbib, floatfix, citeautoscript, amsmath, groupedaddress, aip]{revtex4-1}

\usepackage{fancyhdr}
\fancyheadoffset{0cm}
\usepackage{comment}
\usepackage{bm}
\usepackage{amsmath}
\usepackage{amsfonts}
\usepackage{amssymb}
\usepackage{float}
\usepackage{graphicx}
\usepackage{dcolumn}
\usepackage{mathtools}

\usepackage{hyperref}
\hypersetup{
    colorlinks=true,
    linktocpage=true,
    linkcolor=blue,
    filecolor=magenta,      
    urlcolor=blue,
    citecolor=blue,
    bookmarks=true,
}

\makeatletter
\renewcommand*\l@section{\@dottedtocline{1}{1.5em}{2.3em}}
\makeatother

\graphicspath{{Graphics/}}

\newcommand{\ds}{\displaystyle}

\newcommand{\Q}{{\bf Q}}

\newcommand{\R}{{\bf r}}

\newcommand{\V}{{\bf v}}

\newcommand{\K}{{\bf k}}

\newcommand{\ep}{{\epsilon}}

\newcommand{\be}{\begin}
\newcommand{\en}{\end}
\newcommand{\eq}{equation}

\newcommand{\delep}{\delta\epsilon}
\newcommand{\etal}{\textit{et al.}}
\newcommand{\nubar}{\bar{\nu}}

\begin{document}

\lhead{\small{D. M. Riffe and R. B. Wilson}}
\rhead{\small{Excitation and Relaxation of Nonthermal Electrons}}
\cfoot{--\thepage--}

\title{Excitation and Relaxation of Nonthermal Electron Energy Distributions in Metals with Application to Gold}

\begin{abstract}
A semiempirical theory for the excitation and subsequent relaxation of nonthermal electrons is described.  The theory, which is applicable to ultrafast-laser excited metals, is based on the Boltzmann transport equation for the carrier distribution function $f(\ep,t)$ and includes electron-phonon, electron-electron, and electron-photon scattering integrals in forms that explicitly depend on the electronic density of states.  Electron-phonon coupling is treated by extending the theory of Allen [Phys.~Rev.~Lett.~\textbf{59}, 1460 (1987)] to include highly-excited nonthermal electron distributions, and is used to determine the energy transfer rate between a nonthermal electron subsystem and a thermal phonon subsystem.  Electron-electron scattering is treated with a simple energy-conserving electron-electron scattering integral.  The electron-photon integral assumes photon absorption is phonon assisted.  We apply the theory to analyze prior ultrafast thermionic emission, two-color photoemission, and electronic inelastic light (Raman) scattering experiments on Au.  These analyses show that getting the details of $f(\ep,t)$ is necessary for proper interpretation of each experiment.  Together, the photoemission and Raman-scattering analyses indicate an electron excited 1 eV above the Fermi level has an electron-electron scattering time in the range of 25 to 55 fs. 

\end{abstract}

\author{D. M. Riffe}
\email[Author to whom correspondence should be addressed; electronic mail: ]{mark.riffe@usu.edu}

\affiliation{Physics Department, Utah State University, Logan, UT 84322, USA}

\author{Richard B. Wilson}

\affiliation{Mechanical Engineering Department and Materials Science and Engineering Department, University of California, Riverside, CA 92521, USA}

\date{\today}

\maketitle

\section{Introduction}

Gold has long been a standard for studying ultrafast carrier dynamics in metals.  The earliest experimental work established that ultrafast-laser-pulse excitation of Au rapidly heats the electrons, driving them out of equilibrium with the phonons \cite{Schoenlein1987,Brorson1990,Elsayed-Ali1991}.  By following the re-establishment of equilibrium between electronic and vibrational degrees of freedom, researchers gained insight into the coupling between electrons and phonons.  Key to interpretation of this early data was the use of the two-temperature (2T) model \cite{anisimov1974}, in which the electrons and phonons subsystems are each assumed to be internally equilibrated at some particular (time dependent) temperature, $T_e(t)$ for the electrons and $T_p(t)$ for the phonons.  Equilibrium is purportedly established when $T_e(t) = T_p(t)$.  However, soon after these initial experiments both Groeneveld \textit{et al.} \cite{Groeneveld1992} and Fann \textit{et al.} \cite{Fann1992A,Fann1992} discovered the early-time dynamics are rather more complicated:  it can take significant time -- on the order of several hundred fs -- for the electrons to internally equilibrate and thereby establish a well defined carrier temperature $T_e$.

In order to quantitatively understand their experimental results, Groeneveld \etal~introduced utilization of the Boltzmann transport equation (BTE)  -- with realistic expressions for the electron-phonon and electron-electron scattering integrals -- to the field of ultrafast carrier dynamics in metals \cite{Groeneveld1992,Groeneveld1995}.  With the BTE one can assess the time dependence of the carrier distribution function $f(\ep,t)$ (here $\ep$ is the electron energy), which allows the approach to carrier equilibrium to be followed in detail.  From BTE studies by Groeneveld \textit{et al.} \cite{Groeneveld1992,Groeneveld1995} and others \cite {Sun1994,Tas1994,Gusev1998,Wilson2020} several key facts have been established.  (i) The finite time to establish intracarrier equilibrium inhibits energy transfer from the electron subsystem to the phonon subsystem.  Energy is thus retained by the carriers for longer than it would be otherwise.  (ii) At early times $f(\ep,t)$ is weighted to higher energies than an equivalent-energy thermal distribution.  A significant fraction of hot carriers thus persists for a finite amount of time.  (iii) Relaxation of carriers at the energy scale of an eV is typically controlled by electron-electron scattering.  All of these features are missing from the 2T model, owing to its assumption of instantaneous intracarrier thermalization. 

Details of $f(\ep,t)$ in plasmonic Au nanostructures \cite{Brongersma2015,Hartland2017,Mateo2021} are relevant to several fields of active investigation, including photocatalysis \cite{Linic2015,Zhang2017,Aslam2018,Szczerbinski2018}, solar-energy conversion \cite{Clavero2014,Wu2015,Jang2016}, and inelastic light scattering \cite{Szczerbinski2018,Huag2015,Roloff2017,Barella2021,Xie2016,Jones2018,Carattino2018,Crampton2018,Baffou2021,Jollans2020,Hogan2020,Wu2021,Huang2014,Hugall2015,Cai2019}, the latter of which is used for both nanoscale imaging \cite{Huag2015,Roloff2017,Barella2021} and thermometry \cite{Szczerbinski2018,Xie2016,Jones2018,Carattino2018,Crampton2018,Jollans2020,Barella2021,Baffou2021}.  A number of these light-scattering studies reveal spectra with apparent contribution from nonthermal carriers \cite{Szczerbinski2018,Huang2014,Hugall2015,Jollans2020,Hogan2020,Wu2021}, which supports the fairly widespread notion that out-of-equilibrium carriers are important to photocatalysis and energy conversion in these systems.

However, this notion is not universally held.  Dubi and Sivan have argued that distinguishing the contribution of hot carriers from that of localized heating in catalytic plasmonic systems is not straightforward \cite{Sivan2019a,Dubi2020,Sivan2020b,Sivan2020,Dubi2022}.  Towards the goal of disentangling hot-carrier and heating effects, Dubi and Sivan have put forward several theoretical approaches to calculating electron distributions in nanoplasmonic systems \cite{Sivan2019b,Dubi2019,Sivan2021}, including one that is based on the BTE \cite{Dubi2019}.  However, even in that approach they employ a simple relaxation-time approximation (RTA) for the electron-electron scattering integral.

Several other recent experiments on plasmonic Au systems directly reveal a nonthermal character to the carrier distribution function upon laser excitation \cite{Heilpern2018,Budai2022,Reddy2020}.  Unfortunately, theoretical modeling of these data is either absent \cite{Heilpern2018,Budai2022} or is simplified, owing to the use of the RTA for both electron-phonon and electron-electron scattering integrals \cite{Reddy2020}. 

Overall, use of the BTE for experimental analysis is rather sparse.  Indeed, we have been able to locate only a small number of additional studies in which the BTE has been used to model nonequilibrium-timescale data in metals:  pump-probe optical spectroscopy of Au and Ag films \cite{Suarez1995,Fatti2000}, picosecond ultrasonics in Al \cite{Tas1994}, time-resolved two-photon photoemission (TR2PPE) from Cu \cite{Knorren2001,Knorren2001B} and ferromagnetic transition metals \cite{Knorren2000}, fs pump-probe reflectivity from heavy-fermion compounds LuAgCu$_4$ and YbAgCu$_4$ \cite{Demsar2003,Ahn2004}, fs photoemission from supported Ag nanoparticles \cite{Pfeiffer2004} and graphite \cite{Na2020}, time resolved absorption by colloidal Au nanoparticles \cite{Brown2017}, ultrafast photoluminescence from Ag \cite{Ono2020}, and ultrafast demagnetization in Co/Pt multilayers \cite{Wilson2017}.

This state of affairs motivates the work presented here.  First, we carefully derive a consistent set of BTE scattering integrals that describe the dynamics of a laser excited carrier distribution. Starting with $k$-space-sum descriptions of the integrals we obtain simpler energy integrated forms.  We do this not only for the electron-phonon and electron-electron integrals, but also for an electron-photon integral that describes indirect (phonon mediated) photon absorption.  For all scattering integrals we then derive simplified equations by assuming constant matrix elements.  However, in our final expressions we retain dependence upon the density-of-states (DOS) function $g(\ep)$.  The explicit presence of $g(\ep)$ provides for accurate application of the theory to experimental data on a wide variety of metals.  For the electron-phonon integral we take advantage of the relatively small phonon energy to further derive a differential expression for that term.  In contrast to some approaches to simplifying the BTE \cite{Gusev1998,Kabanov2008,Wilson2020}, we do not linearize the scattering integrals in the deviation from the underlying FD function; our approach is thus applicable to high excitation levels that are often present in experimental studies.

Second, we demonstrate the utility of our BTE theory by applying it to three distinct experimental studies of Au:  femtosecond thermionic emission \cite{wang1994}, two-color time-resolved photoemission \cite{Fann1992}, and subpicosecond inelastic light (Raman) scattering from a plasmonic system \cite{Huang2014}.  New insight into all three experiments is obtained.  In the case of the thermionic-emission experiment we illustrate the inadequacy of the (previously utilized) 2T model in describing those data.  From analysis of the photoemission data we assess the strength of carrier-carrier scattering.  Our analysis of the Raman scattering data demonstrates the highly nonthermal nature of the carrier distribution in that experiment.   Together, our analysis of all three experiments highlights the necessity of accurate assessment of the time dependent carrier distribution $f(\ep,t)$ at fs timescales.

\section{The BTE Model}
\label{SecII}

The starting point for the development of our theory is the BTE for the electron distribution function $f_k(\R,t)$.  For near-ir or visible excitation of a metal by an ultrafast laser pulse the BTE can be written as
\be{\eq}
\label{1}
\frac{\partial f_k}{\partial t} + \V_k \cdot \nabla_r f_k = \bigg( \! \frac{\partial f_k}{\partial t} \!  \bigg)_{\!\! s}.
\en{\eq}
Here $\R$ represents real-space coordinates, $k = \{ \K,n \}$ represent the combination of electron wave $\K$ and band index $n$, $\V_k$ is the velocity of an electron labeled by $k$, and the term on the right-hand side accounts for changes in the distribution function due to electronic scattering processes.  In this paper we assume spatially uniform laser excitation of the solid.  The spatial dependence of $f$ and, thus, the transport term on the left side of Eq. (\ref{1}) can be neglected.  This approximation is applicable to the case of a laser-excited thin film with a thickness $d_s$ comparable to or less than the optical skin depth $\delta$ of the material.  For typical metals excited by laser pulses in the near ir or visible $\delta \sim$ 10 nm.  Treatment of thicker films or bulk samples requires that electron transport be included in the time evolution of the hot electrons \cite{Gusev1998}.

For the right-hand side of Eq.~(\ref{1}) we include three scattering processes: electron-phonon, electron-electron, and electron-photon so that
\be{\eq}
\label{2}
\bigg(\! \frac{\partial f_k}{\partial t} \! \bigg)_{\!\! s} = \bigg(\! \frac{\partial f_k}{\partial t} \! \bigg)_{\!\! ep} + \bigg( \! \frac{\partial f_k}{\partial t}  \!\bigg)_{\!\! ee} + \bigg( \! \frac{\partial f_k}{\partial t} \! \bigg)_{\!\! e\gamma}
\end{\eq}
For a nonthermal distribution electron-electron scattering provides the means for the electrons to thermalize among themselves.  Electron-phonon scattering is the energy exchange mechanism between the hot electrons and the (generally) cooler phonon subsystem.  The electron-photon scattering term is used to describe excitation of the electrons by the laser pulse.  We note that we ignore scattering by impurities.  Since such scattering is elastic or quasielastic, it mainly serves to randomize the electron distribution in $k$ space.  However, since we are treating a uniformly excited solid, impurity scattering will have no effect on the relaxation of the distribution function $f_k$.\footnote{Insofar as impurity scattering may enhance electron-phonon coupling, it can be included in the electron-phonon scattering term by adjustment of the strength of that term.}

Our general approach to treating the three scattering terms in Eq.~(\ref{2}) is to start with a Fermi-golden-rule expression that describes the time-dependence of the change in $f_k$.  Because elastic and/or nearly-elastic scattering processes are extremely rapid (typical scattering times are $\sim$ 10 fs at room temperature), the distribution function is rapidly randomized in $k$ space.  This randomization justifies the use of a distribution function $f(\ep,t)$ that depends upon $k$ only through the electron energy $\ep$.  Concurrently, we introduce $k$-space averaged scattering strengths that transform Eq.~(2) into an equation for $\partial f(\ep) / \partial t$.  With this approach the dependence of the three scattering terms in Eq.~(\ref{2}) on the electronic density of states (DOS) $g(\ep)$ is clearly revealed.  In comparison to many other models of electron dynamics, we do not assume $g(\ep)$ is constant.  Our approach to treating these scattering terms is very much in the spirit of the random-$k$ approximations of Berglund and Spicer \cite{berglund1964} and Kane \cite{kane1967}.

As we discuss in more detail in Secs.~\ref{Sec11.A} and \ref{Sec11.B}, our transformation of the BTE and relevant scattering integrals [Eqs.~(\ref{1}) and (\ref{2})] from $k$-space to energy space is physically justified for our study of electron dynamics in Au.  From a practical point of view, this transformation is also computationally necessary.  In principle, scattering strengths in Eq.~(\ref{2}) could be populated from first principles without $k$-space averaging. Computational methods based on density functional theory make it possible to predict how electron-phonon \cite{Bernardi2015,Mustafa2016}, electron-electron \cite{Zhukov2002a,Zhukov2002b,Ladstadter2004,Bernardi2015}, and electron-photon \cite{Bernardi2015,brown2016a} interaction strengths vary across momentum space.  However, the nature of scattering processes in Eq.~(\ref{2}) requires a large number of $k$-space points to be evaluated for convergence.  Electron-electron and electron-photon interactions couple states with energy differences on the order of the photon energy $h \nu$, while the electron-phonon interaction couples states with energy differences on the order of phonon energies $\hbar \omega$.  Therefore, a converged solution of the BTE requires sampling $N_s \approx (\nu / \omega)^3$ $k$-space points within the Brillouin zone. In practical terms, in reciprocal space Eqs.~(\ref{1}) and (\ref{2}) represent somewhere between $10^6$ and $10^9$ coupled nonlinear integro-differential equations, each with $10^6$ to $10^9$ scattering events \cite{Bernardi2022}.  These equations need to be solved across ps timescales with fs timescale increments (i.e., with $\sim 10^3$ to $10^4$ time steps). Methods like Monte Carlo sampling \cite{brown2016a} and/or Wannier interpolation \cite{Marzari1997} can be used to reduce the computational burden. However, even when such methods are implemented for problems much simpler than solving the BTE, such as electrical-resistivity calculations for simple metals, more than $10^6$ reciprocal-space points are needed for convergence \cite{brown2016a}.  In 2D-materials, the reduced dimensionality allows convergence with only $N_s \approx (\nu / \omega)^2$. This has allowed several recent studies to compute solutions to the BTE in reciprocal space for graphene \cite{Bernardi2022,Wadgaonkar2022}.  However, to our knowledge, no similar studies currently exist for metals like Au.  An alternative first-principles method for modeling hot electron dynamics is real-time time-dependent perturbation theory.  While this approach has been successful for modeling ultrafast electron dynamics in molecules \cite{Andrea2013}, this technique has so far struggled to model hot electron dynamics in metals \cite{Silaeva2018,Kononov2022}.

\bigskip

\begin{widetext}
\subsection{Electron-phonon scattering}
\label{Sec11.A}

For the electron-phonon scattering term in Eq.~(\ref{2}) we start with the expression of Lifshitz and Pitaevskii (LP) \cite{LLPK}, but  expressed -- for the most part -- in the notation of Allen \cite{allen1987},
\begin{align}
\label{3}
\bigg( \! \frac{\partial f_{k}}{\partial t} \! \bigg)_{\!\! ep} = - \frac{2 \pi }{\hbar} \sum_{k',Q} &\Big\{ \big| M_{k'Q,k}^{ep}  \big|^2  \delta_{\K,\K'+\Q} \delta(\ep_k - \ep_{k'} - \hbar \omega_Q)  \big[  f_k (1-f_{k'}) (N_Q + 1) - (1-f_k) f_{k'} N_Q \big] \nonumber \\
		& +\big| M_{k',kQ}^{ep}  \big|^2 \delta_{\K',\K+\Q} \delta(\ep_k - \ep_{k'} + \hbar \omega_Q) \big[   f_k (1-f_{k'}) N_Q - (1-f_k) f_{k'} (N_Q + 1) \big] \Big\}.
\end{align}
Here $Q = \{ \Q,j \}$ represents the combination of the phonon wave vector $\Q$ and mode index $j$, and $N_Q$ is the phonon distribution function.  The quantity $M_{k'Q,k}^{ep}$ ($M_{k',kQ}^{ep}$) is the matrix element for scattering an electron from state $k$ to state $k'$ with the emission (absorption) of a phonon described by $Q$.  The equality $\big| M_{k,k'Q}^{ep}  \big| =\big| M_{k'Q,k}^{ep}  \big|$, which derives from these two matrix elements describing processes that are the inverse of each other, has been utilized in writing Eq.~(\ref{3}).  We assume all wave vectors $\K$, $\K'$, and $\Q$ are in the first Brillouin zone; a consequence of this assumption is that an implied reciprocal lattice vector is present in the Kronecker delta functions $\delta_{\K,\K'+\Q}$ and $\delta_{\K',\K+\Q}$.

We now take advantage of an approximate equality noted by LP,
\be{\eq}
\label{3b}
\big| M_{k'Q,k}^{ep}  \big|^2 \approx \big| M_{k',k(-Q)}^{ep}  \big|^2.
\en{\eq}
That is, the probability of scattering an electron from state $k$ to state $k'$ is largely independent of whether a phonon is absorbed or emitted in the process.  We note this near equality follows from $\hbar \omega_Q \ll \ep_F $, where $\ep_F$ is the Fermi energy.  If we now utilize Eq.~(\ref{3b}), assume lattice symmetry implies $\hbar \omega_{-Q} = \hbar \omega_{Q}$, and also assume $N_{-Q} = N_Q$, then we can then readily simplify Eq.~(\ref{3}) to 
\begin{align}
\label{3c}
\bigg( \! \frac{\partial f_{k}}{\partial t} \! \bigg)_{\!\! ep} = - \frac{2 \pi }{\hbar} \sum_{k',Q} & \big| M_{k' \! ,k}^{ep}  \big|^2  \delta_{\K,\K'+\Q} \, \Big\{  \delta(\ep_k - \ep_{k'} - \hbar \omega_Q)  \big[  f_k (1-f_{k'}) (N_Q + 1) - (1-f_k) f_{k'} N_Q \big] \nonumber \\
		& +  \delta(\ep_k - \ep_{k'} + \hbar \omega_Q) \big[   f_k (1-f_{k'}) N_Q - (1-f_k) f_{k'} (N_Q + 1) \big] \Big\}.
\end{align}
Here we have also taken advantage of the near independence from $Q$ to write the matrix element as only depending on $k$ and $k'$.  Aside from (i) a difference in the normalization of the matrix element $M_{k' \!,k}$ and (ii) our explicit inclusion of $\delta_{\K,\K+\Q}$, Eq.~(\ref{3c}) is equivalent to Eq.~(1) of Allen \cite{allen1987}.

To proceed to our $k$-space averaged expression we multiply both sides of Eq.~(\ref{3c}) by
$$
\delta(\ep - \ep_k) \, \delta(\ep' - \ep_{k'}) \, \delta(\Omega - \omega_Q) \, \frac{g^{\uparrow}(\ep_k)}{g^{\uparrow}(\ep)}
$$
and integrate on the variables $\ep_k$, $\ep'$, and $\Omega$.  Here
\be{\eq}
\label{4}
g^{\uparrow}(\ep) = \sum_k \delta(\ep - \ep_k)
\en{\eq}
is the spin-resolved density of electronic states.\footnote{Because we are only considering nonmagnetic materials, $g^{\uparrow}(E) = g^{\downarrow}(E) = g(E)/2$.}  If we assume the electron-distribution function $f_k$ depends upon $k$ only through the energy $\ep_k$ [i.e., $f_k = f(\ep_k)$; this is the essence of the random-$k$ approximation] and likewise assume $N_Q$ only depends upon $Q$ via $\omega_Q$, then we obtain
\begin{align}
\label{5}
\bigg( \! \frac{\partial f(\ep)}{\partial t} \! \bigg)_{\!\! ep} = - 2 \pi  \frac{g^{\uparrow}(\ep_F)}{g^{\uparrow}(\ep)} &\int d\Omega \int d\ep' \, \alpha^2 F(\ep,\ep',\Omega) \Big\{ \delta(\ep - \ep' - \hbar \Omega)  \nonumber \\
		&\times \big[  f(\ep) (1-f(\ep')) (N(\Omega) + 1) - (1 - f(\ep)) f(\ep') N(\Omega) \big] \nonumber \\
		& + \delta(\ep - \ep' + \hbar \Omega) \big[   f(\ep) (1-f(\ep')) N(\Omega) - (1-f(\ep)) f(\ep') (N(\Omega) + 1) \big] \Big\} ,
\end{align}
where
\be{\eq}
\label{6}
\alpha^2 F(\ep,\ep',\Omega) = \frac{1}{\hbar \, g^{\uparrow}(\ep_F)} \sum_{k,k',Q} \big| M_{k' \!,k}^{ep}  \big|^2 \delta(\ep - \ep_k) \delta(\ep' - \ep_{k'}) \delta(\Omega - \omega_Q) \, \delta_{\K,\K'+\Q}
\en{\eq}
is the electron-phonon spectral density function.  This function can be interpreted as a $k$-space averaged electron-phonon scattering strength.  Since we are interested in cases where electrons are excited far above the Fermi level (and holes far below the Fermi level) we do not make the usual approximation for $\alpha^2F$ that $\ep$ and $\ep'$ are equal to $\ep_F$.
We note the derivation of Eqs.~(\ref{5}) and (\ref{6}) takes advantage of the relation
\be{\eq}
\label{6b}
\int g^\uparrow(\ep_k) \delta(\ep - \ep_k) f(\ep_k) \, d\ep_k = \sum_k \delta(\ep - \ep_k) f_k
\en{\eq}
to express the integration over $\ep_k$ of the right side of Eq.~(\ref{3c}) as a sum over $k$.  

While Eq.~(\ref{5}) is the basic result of our $k$-space averaged approach, there are several additional, useful approximations that can be made.  First, we assume $\big| M_{k' \!,k}^{ep}  \big|$ is constant for all scattering events.  With $\big| M_{k' \!,k}^{ep}  \big|$ constant Eqs.~(\ref{4}) and (\ref{6}) imply \cite{wang1994}
\be{\eq}
\label{7}
\alpha^2 F(\ep,\ep',\Omega) = \frac{g(\ep) \, g(\ep')}{g^2(\ep_F)} \, \alpha^2F(\Omega),
\en{\eq}
where $\alpha^2F(\Omega) = \alpha^2 F(\ep_F,\ep_F,\Omega)$.  
If we insert this relation in Eq.~(\ref{5}) and evaluate the $\ep'$ integral, then we find
\begin{align}
\label{6c}
\bigg( \! \frac{\partial f(\ep)}{\partial t} \! \bigg)_{\!\! ep} = - 2 \pi   \int d\Omega \,\, \alpha^2 F(\Omega) &\bigg\{ \frac{g(\ep - \hbar\Omega)}{g(\ep_F)} \big[ f(\ep)(N(\Omega) + 1) - f(\ep) f(\ep - \hbar \Omega) - f(\ep - \hbar \Omega) N(\Omega) \big] \nonumber \\
		& + \frac{g(\ep + \hbar\Omega)}{g(\ep_F)} \big[ f(\ep)N(\Omega) + f(\ep) f(\ep + \hbar \Omega) - f(\ep + \hbar \Omega) (N(\Omega) + 1) \big] \bigg\} .
\end{align}
To proceed further we (i) assume $N(\Omega)$ is thermal is nature and thus described by the phonon temperature $T_p$\footnote{The assumption of a thermal lattice at femtosecond timescales is a rough approximation.  However, treating phonons with the same level of detail that we treat electrons is well outside the scope of this paper, and so we shall proceed with this simplification.  Furthermore, because our primary interest is in relatively low excitation levels, we simply assume a constant phonon temperature $T_p$ in all of our calculations.} and (ii) thence expand $g(\ep \pm \hbar \Omega)$, $f(\ep \pm \hbar \Omega)$, and $N(\Omega) = (e^{\hbar \Omega / k_BT_p} -1)^{-1}$ in a Taylor's series in powers of $\hbar \Omega$.  This expansion produces, to linear order in $\hbar \Omega$,
\be{\eq}
\label{8}
\bigg( \! \frac{\partial f(\ep)}{\partial t} \! \bigg)_{\!\! ep}^{\! \!(\Omega)} = \frac{\pi \hbar \, \lambda \langle \Omega^2 \rangle}{g(\ep_F)} \Big\{ g(\ep) \Big[  f''(\ep) k_B T_p + f'(\ep) \big[ 1 - 2f(\ep) \big]   \Big]  + 2g'(\ep) \Big[ f'(\ep)k_B T_p + f(\ep) \big[  1 - f(\ep) \big]  \Big] \Big\},
\en{\eq}
where the superscript $(\Omega)$ signifies the right side of this equation is the leading (linear) term in the expansion.  Here $f'(\ep) = \partial f / \partial \ep$, e.g., and we have used the standard definition
\be{\eq}
\label{9}
\lambda \langle \Omega^n \rangle = 2 \int d\Omega \, \alpha^2F(\Omega) \, \Omega^{n-1},
\en{\eq}
where $\lambda = \lambda \langle \Omega^0 \rangle$ is the is the mass enhancement factor from superconductivity theory.  As discussed in Appendix \ref{Appendix B} and utilized below, $\langle \Omega^n \rangle$ can be estimated using moment Debye temperatures.  For $k_B T_p \gtrsim \hbar \Omega$ higher-order terms in $\hbar \Omega$ are unnecessary.  This is discussed in more detail in Appendix \ref{Appendix A}.  An appealing feature of Eq.~(\ref{8}) is that it only depends upon two material-dependent quantities:  the electron-phonon scattering strength $\lambda \langle \Omega^2 \rangle$ and the electronic density of states $g(\ep)$.  We note in passing that under conditions of constant $g(\ep)$ Eq.~(\ref{8}) reduces to Eq.~(6) of Ref.~[\onlinecite{Gusev1998}].

The rate of energy transfer from the electrons to the phonons is given by
\be{\eq}
\label{10}
\bigg( \! \frac{d \langle \ep \rangle }{dt}  \! \bigg)_{\!\! ep} = \int d\ep \, g(\ep) \, \ep \, \bigg( \! \frac{\partial f(\ep)}{\partial t}  \! \bigg)_{\!\! ep}.
\en{\eq}
Noting Eq.~(\ref{8}) can be rewritten as 
\be{\eq}
\label{8b}
\bigg( \! \frac{\partial f(\ep)}{\partial t} \! \bigg)_{\!\! ep}^{\! \!(\Omega)} = \frac{\pi \hbar \, \lambda \langle \Omega^2 \rangle}{g(\ep_F) \, g(\ep)} \, \frac{d}{d\ep} \Big\{ g^2(\ep) \Big[ f'(\ep)k_B T_p + f(\ep) \big[  1 - f(\ep) \big]  \Big] \Big\},
\en{\eq}
we use Eq.~(\ref{8b}) in Eq.~(\ref{10}) which (after an integration by parts) results in 
\be{\eq}
\label{11}
\bigg( \! \frac{d \langle \ep \rangle }{dt}  \! \bigg)_{\!\! ep}^{\! \!(\Omega)} = \frac{\pi \hbar \, \lambda \langle \Omega^2 \rangle}{g(\ep_F)} \int d\ep \, g^2(\ep) \big\{ -f'(\ep) k_B T_p - f(\ep) [ 1 - f(\ep)  ]  \big\}.
\en{\eq}
We note Eqs.~(\ref{8}) and (\ref{11}) are the key results of our theory with regard to electron-phonon scattering.

\end{widetext}

Connection to 2T model \cite{anisimov1974} of electron-phonon dynamics is made via Eq.~(\ref{11}) with the assumption $f(\ep)$ is a FD distribution described by an electron temperature $T_e$.  In that case $f(\ep)[1 - f(\ep)] = -f'(\ep) k_BT_e$ and Eq.~(\ref{11}) becomes
\be{\eq}
\label{12}
\frac{1}{V}\bigg( \! \frac{d \langle \ep \rangle_{\scriptscriptstyle \! F\!D} }{dt}  \! \bigg)_{\!\! ep}^{\! \!(\Omega)} = G_{ep}(T_e) \, (T_p - T_e),
\en{\eq}
where  $\langle \ep \rangle_{\scriptscriptstyle \! F\!D}$ is the excess energy the thermal distribution, and 
\be{\eq}
\label{13}
G_{ep}(T_e) = \frac{\pi \hbar  \lambda \langle \Omega^2 \rangle k_B}{g_v(\ep_F)} \int d\ep \, g_v^2(\ep) [- f'_{\scriptscriptstyle \! F\!D}(\ep)]
\en{\eq}
is a  temperature dependent electron-phonon coupling parameter.  Here $V$ is the volume of the solid, and $g_v = g/V$ is the volume normalized DOS [J$^{-1}$m$^{-3}$].  As is evident from Eq.~(\ref{13}), $G_{ep}$ is constant for relatively low electron temperatures, in which case $-f'_{\scriptscriptstyle \! F\!D}(\ep) \approx \delta(\ep - \ep_F)$, and the canonical electron-phonon coupling constant\cite{allen1987}
\be{\eq}
\label{12b}
G_{ep} = \pi \hbar  \, k_B \lambda \langle \Omega^2 \rangle \, g_v(\ep_F)
\en{\eq}
is obtained.  Equations (\ref{12}) and (\ref{13}) were previously derived and then used to analyze femtosecond thermionic emission data, where it was assumed the 2T model was appropriate \cite{wang1994}.  In Sec.~\ref{SecV.B} below we apply the results of the present BTE model to investigate the validity of the 2T-model assumption in the experimental regime germane to that study.  We note Lin \textit{et al}. \cite{Lin2008} have outlined a similar derivation of Eqs.~(\ref{12}) and (\ref{13}).

\begin{figure}[b!]
\centerline{\includegraphics[scale=0.60]{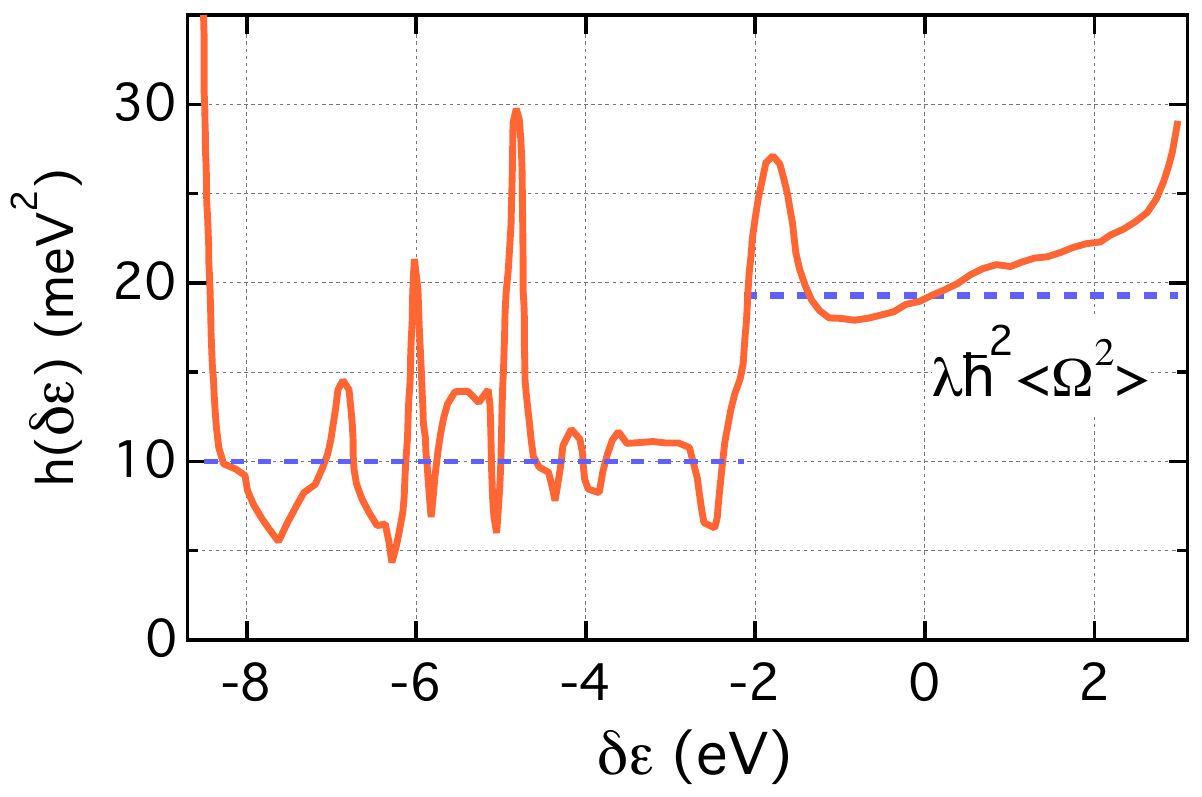}}
\caption{Energy dependent electron-scattering strength $h(\delta \ep)$ for Au from Brown \textit{et al} \cite{Brown2016b}.  Here $h(\delta \ep)$ has been scaled (by 0.70) to match $\lambda \hbar^2 \langle \Omega^2 \rangle$ = 19.3 meV$^2$ at $\ep = \ep_F$.}
\label{Fig12}
\end{figure}

How valid is the constant $\big| M_{k' \!,k}^{ep}  \big|$ approximation?  Brown \etal \cite{Brown2016b} address this issue with first-principles calculations of an electron-phonon coupling-strength function $h(\delep)$ for Al, Cu, Ag, and Au (here $\delep = \ep - \ep_F$ is the electron energy $\ep$ referenced to the Fermi energy $\ep_F$).  We note if $h(\delep)$ were constant, then it would simply equal $\lambda \hbar^2 \langle \Omega^2 \rangle$.  For Al the function $h(\delep)$ monotonically decreases vs $\delep$, and the decrease is slow enough that letting $h(\delep)$ be constant over a range of a few eV about $\delep = 0$ is an excellent approximation.  For the noble metals the situation is rather more complicated, owing to the $d$-electron band that that resides not too far below $\ep_F$.  This is illustrated in Fig.~\ref{Fig12} , where we plot $h(\delep)$ for Au.    Because of the equality mentioned above, in the figure we have scaled $h(\delep)$ so that $h(0) = \lambda \hbar^2 \langle \Omega^2 \rangle = 19.3$ eV$^2$ (see Sec.~\ref{SECIII} for a discussion of $\lambda \langle \Omega^2 \rangle$ for Au).  While there is ample structure in $h(\delep)$, overall a distinct division in behavior exists between the $sp$ and $d$ electrons.  For the $sp$ band ($\delep \gtrsim - 2$ eV) $h(\delep)$ is (i) never much less than $h(0)$ and (ii) typically not greater than $\sim$ 30\% above $h(0)$.  A decent approximation in this range is $h(\delep) \approx 21$ meV$^2$.  Conversely, in the $d$-band range of energies ($\delep \lesssim - 2$ eV) a reasonable approximation for $h(\delep)$ is 10 meV$^2$ (shown as the dashed line at lower energies), about half the average value for the $sp$ band.  The behavior of $h(\delep)$ for Cu and Ag is similar to that for Au  \cite{Brown2016b}.  For the noble metals we thus conclude the constant $\big| M_{k' \!,k}^{ep}  \big|$ approximation is reasonable if the $d$ electrons do not get involved in the dynamics.

\begin{widetext}
\subsection{Electron-electron scattering}
\label{Sec11.B}

Our starting point for the electron-electron scattering contribution in Eq.~(\ref{2}) is Eq.~(A2) from Penn, Apell, and Girvin \cite{penn1985}, from which we obtain
\begin{align}
\label{14}
\bigg( \! \frac{\partial f_{k}}{\partial t} \! \bigg)_{\!\! ee} = - \frac{4 \pi}{\hbar} \sum_{k_2,k_3,k_4} &\Big\{ \big| M^{ee}_{k_3k_4,k k_2}  \big|^2   \delta_{\K+\K_2,\K_3 + \K_4} \delta(\ep_k +\ep_{k_2} - \ep_{k_3} - \ep_{k_4}) \nonumber \\
	& \times  \big[  f_k f_{k_2} (1 - f_{k_3}) (1 - f_{k_4} ) -  (1-f_k)(1-f_{k_2}) f_{k_3} f_{k_4}  \big] \Big\}.
\end{align}
Equation (\ref{14}) includes not only scattering out of the state $k$ (as in the case of Eq.~(A2) of Ref.~[\onlinecite{penn1985}]) but also scattering into the state $k$.  In Eq.~(\ref{14}) the spin sums have already been done and interference between direct and exchange scattering is assumed to be negligible.  Calculations by Penn, Apell, and Girvin justify the neglect of this interference in cases where electron-electron scattering transitions are averaged over $k$-space \cite{penn1985}.  The matrix element $M^{ee}_{k_3k_4,k k_2}$ is for non-spin-flip scattering of two electrons from states $k$ and $k_2$ to states $k_3$ and $k_4$; we assume this matrix element is independent of the spin states of the electrons.  In writing Eq.~(\ref{14}) we have taken advantage of the equality $\big| M^{ee}_{k k_2, k_3 k_4}  \big|^2 = \big| M^{ee}_{k_3 k_4, k k_2}  \big|^2$.  As with the basic integral for electron-phonon scattering  [Eq.~(\ref{3})], we assume the wave vectors (in this case $\K$, $\K_2$, $\K_3$, and $\K_4$) all lie within the first Brillouin zone, with the similar consequence that an implied reciprocal vector is present in the Kronecker delta function $\delta_{\K+\K_2,\K_3 + \K_4}$ \cite{kane1967}.

Following a formalism much the same as for electron-phonon scattering, we multiply both sides of Eq.~(\ref{14}) by
$$
\delta(\ep - \ep_k) \, \delta(\ep_2 - \ep_{k_2}) \, \delta(\ep_3 - \ep_{k_3}) \delta(\ep_4 - \ep_{k_4}) \, \frac{g^{\uparrow}(\ep_k)}{g^{\uparrow}(\ep)}
$$
and integrate on the variables $\ep_k$, $\ep_2$, $\ep_3$, and $\ep_4$ [again making use of Eq.~(\ref{6b})].  As above, with the assumption the electron distribution is a function only of energy (and not momentum or band index) we obtain
\begin{align}
\label{15}
\bigg( \! \frac{\partial f(\ep)}{\partial t} \! \bigg)_{\!\! ee} &= - \frac{\pi}{2 \hbar} \int d\ep_3 \int d\ep_4 \, M^2_{ee}(\ep_3,\ep_4;\ep,\ep_3 + \ep_4 - \ep) \, \Big( g(\ep_3 + \ep_4 - \ep) g(\ep_3)  g(\ep_4)  \nonumber \\
	&  \times \big\{ f(\ep) f(\ep_3 + \ep_4 - \ep) [1 - f(\ep_3)] [1- f(\ep_4)]  - [1 - f(\ep)] [1-f(\ep_3 + \ep_4 - \ep)] f(\ep_3) f(\ep_4)  \big\} \Big)
\end{align}
where
\be{\eq}
\label{16}
M^2_{ee}(\ep_3,\ep_4;\ep,\ep_2) = \frac{ \ds{ \sum_{k,k_2,k_3,k_4} \big| M^{ee}_{k_3k_4,k k_2}  \big|^2 \delta(\ep - \ep_k) \, \delta(\ep_2 - \ep_{k_2}) \, \delta(\ep_3 - \ep_{k_3}) \delta(\ep_4 - \ep_{k_4})  \delta_{\K+\K_2,\K_3+\K_4} }}{ \ds{\sum_{k,k_2,k_3,k_4}  \delta(\ep - \ep_k) \, \delta(\ep_2 - \ep_{k_2}) \, \delta(\ep_3 - \ep_{k_3}) \delta(\ep_4 - \ep_{k_4})  }}.
\en{\eq}
Notice $M^2_{ee}(\ep_3,\ep_4;\ep,\ep_2)$ is the average of $\big| M^{ee}_{k_3k_4,k k_2}  \big|^2$ over the four energy surfaces $\ep_3$, $\ep_4$, $\ep$, and $\ep_2$.  The numerator in Eq.~(\ref{16}) is analogous to $\alpha^2F(\ep,\ep',\Omega)$ for electron-phonon scattering.  We now make the approximation $M^2_{ee}(\ep_3,\ep_4;\ep,\ep_2)$ is constant.  As discussed in detail below, this approximation appears to be reasonable on the energy scale of at least a few eV.  We thus write
\begin{align}
\label{17}
\bigg( \! \frac{\partial f(\ep)}{\partial t} \! \bigg)_{\!\! ee} &= - \frac{K_{ee}}{g^3(\ep_F)} \int d\ep_3 \int d\ep_4 \Big(  g(\ep_3 + \ep_4 - \ep) g(\ep_3)  g(\ep_4)    \nonumber \\
	& \times \big\{ f(\ep) f(\ep_3 + \ep_4 - \ep) [1 - f(\ep_3)] [1- f(\ep_4)]  - [1 - f(\ep)] [1-f(\ep_3 + \ep_4 - \ep)] f(\ep_3) f(\ep_4)  \big\} \Big),
\end{align}
\end{widetext}
where
\be{\eq}
\label{18}
K_{ee} = \frac{\pi}{2 \hbar} \, M_{ee}^2 \, g^3(\ep_F).
\en{\eq}
Equation (\ref{17}) is the basic equation we use for numerical calculations of electron-electron scattering.  It is analogous to Eq.~(\ref{8}) for electron-phonon scattering in that it explicitly reveals the dependence on the electronic DOS $g(\ep)$.  We note our development here is similar to that of Knorren \textit{et al.} \cite{Knorren2000}, and the equivalent of Eq.~(\ref{17}) has been used by several other researchers to calculate electron-electron scattering induced changes in $f(\ep,t)$ \cite{knorren1999,Ahn2004,Brown2017,Ono2020,Na2020}.

The scattering strength parameter $K_{ee}$ can be obtained in several ways.  It can be calculated using density functional theory, as discussed in Sec.~\ref{SECIII} below.  Conversely, it can be fit to experimental data, as we do in Secs.~\ref{SecVIB} and \ref{SecRaman}.  For simple metals it can also be connected to the Fermi-liquid-theory result of Pines and Nozieres for the energy dependent lifetime $\tau_{ee}(\delep)$ of an electron \cite{Pines1966}.  Indeed, for $\delep \gg k_B T_e$ the scattering strength and this lifetime are related via \cite{Kabanov2008}
\begin{equation}
\label{29}
\frac{1}{\tau_{ee}(\delep)} = \frac{K_{ee}}{2} (\delep)^2.
\end{equation}

It is natural to ask, under what conditions is a constant average matrix element $M_{ee}(\ep_3,\ep_4;\ep,\ep_2)$ a reasonable approximation?  Zarate \textit{et al.} directly address this question by reviewing analyses of prior experimental data on Cu, Ag, and several ferromagnetic materials \cite{Zarate1999}.  Their discussion suggests it is not an unreasonable assumption for an energy range of 3 to 4 eV in these materials.  Their discussion also suggests it may be more important to accurately account for DOS effects [as in Eq.~(\ref{17})] than to worry about the details of $M_{ee}$.  Similarly, calculations by Knorren \textit{et al.}~for Cu that assume a constant $M_{ee}$ qualitatively reveal DOS influenced trends in hot-carrier dynamics \cite{Knorren2000}.

Relevant to this discussion, Ogawa \textit{et al.}~have calculated hot-electron lifetimes in Cu with two methods: the first uses detailed band-structure information, while the second essentially uses Eq.~(\ref{17}) \cite{Ogawa1997}.  Their two calculations produce quite similar results for electron-state lifetime $\tau_{ee}$ vs energy $\ep$, including a broad peak that is also observed experimentally. 

Perhaps the most enlightening investigation into the validity of Eq.~(\ref{17}) comes from Zhukov \textit{et al.} \cite{Zhukov2002a,Zhukov2002b}, who compare single-particle lifetimes obtained using (the first half of) this equation with those from \textit{ab initio} LMTO-RPA-GW calculations.  They study the simple metal Al and the 4$d$ metals Nb, Mo, Rh, Pd, and Ag.  From their \textit{ab initio} calculations of $\tau_{ee}(\delep)$ they extract an $\ep_3$ and $\ep_4$ integrated $M^2_{ee}(\ep_3,\ep_4;\ep,\ep_3 + \ep_4 - \ep)$ [see Eq.~(\ref{15})] and find that as a function of $\ep$ this integrated matrix element typically varies quite slowly with $\ep$.  Furthermore, they show that a judicious choice of constant $M_{ee}$ for each metal results in a $\tau_{ee}(\delep)$ curve [obtained from Eq.~(\ref{29})] that matches quite well the curve obtained from the \textit{ab initio} calculations.  Similar results have also been obtained from first-principles calculations by Ladst\"adter \textit{et al.} \cite{Ladstadter2004} and Bernardi \textit{et al.}\cite{Bernardi2015}

There is evidence, however, that a constant $M_{ee}$ is not always sufficient for describing electron-electron interactions.  This appears to be particularly true for materials with transitions that involve both free-electron-like $sp$ bands and more localized $d$ bands \cite{penn1985,Zarate1999,Drouhin2000,Knorren2001}.  Analyses of experimental data carried out by Penn \textit{et al.} \cite{penn1985}, Zarate \textit{et al.} \cite{Zarate1999}, and Drouhin \cite{Drouhin2000} all suggest -- for the most part -- that $|M_{ee}|^2$ decreases as more $d$ electrons are involved in the scattering.  The analyses by Penn \textit{et al.} and Drouhin also show that if one is interested in scattering rates over an energy range of tens of eV, then one must indeed account for the energy dependence of the matrix element that (simultaneously) involves $sp \rightarrow d$ and $d \rightarrow sp$ transitions.

With regard to our analysis of Au data below, the upshot is similar to that for electron-phonon scattering:  if only $sp$ electrons are involved in the dynamics, then we are justified in assuming a constant matrix element.  This is the case if the photon energy is less than the $\sim \,$2.0 eV transition from the top of the 5$d$ band to the Fermi level and the laser intensity is relatively low.  On the other hand, if holes are created in the normally filled 5$d$ band,  then it may be necessary to go beyond the constant-matrix-element approximation in describing the dynamics.


\subsection{Electron-photon scattering}

Electron-photon scattering processes in a solid can be classified into two categories.  The first category is direct interband scattering, where an electron directly absorbs or emits a photon while making a transition between energy states in two different electronic bands.  Since near-ir and visible photons have very little momentum (compared to a typical electron state) the electron states involved in direct transitions are customarily taken to have the same momentum.  The second category is indirect electronic scattering, which involves emission or absorption of a phonon during the electronic transition.  This scattering process is applicable to intraband excitations, which are the only allowed excitations in Au at photon energies $h \nu \lesssim$ 2 eV.  Since a typical phonon energy is much smaller than the energy of a near-ir or visible photon, it is often convenient to ignore the phonon energy (while still considering the phonon momentum) in this scattering process.  In the case of direct electron-photon scattering, the condition of equal momenta in the initial and final electron states precludes any simple formulation in terms of the DOS of the solid, as has been done for electron-phonon and electron-electron scattering above.  However, in the case of indirect electron-photon scattering we can proceed with a $k$-space averaged formulation, as follows.

For indirect electron-photon scattering the laser-excitation term in Eq. (\ref{2}) can be written as
\begin{widetext}
\begin{align}
\label{19}
\bigg( \! \frac{\partial f_k}{\partial t} \! \bigg)_{\!\! e\gamma} =& - \frac{2 \pi}{\hbar} n_{h\nu}(t) \sum_{k',Q} \bigg\{ \delta(\ep_k - \ep_{k'} + h \nu) f_k (1 - f_{k'}) \Big[ \big| M_{k' Q,k}^{e \gamma}  \big|^2 (N_Q \! + \! 1) \delta_{\K,\K' + \Q} +  \big| M_{k', kQ}^{e \gamma}  \big|^2 N_Q \delta_{\K',\K + \Q} \Big] \nonumber \\
	& - \delta(\ep_k - \ep_{k'} - h \nu) f_{k'} (1 - f_k) \Big[ \big| M_{k Q,k'}^{e \gamma}  \big|^2 (N_Q \! + \! 1) \delta_{\K',\K + \Q} +  \big| M_{k, k'Q}^{e \gamma}  \big|^2 N_Q \delta_{\K,\K' + \Q} \Big] \bigg\},
\end{align}
where $n_{h \nu}(t)$ is proportional to the time-dependent intensity of the laser pulse, which is centered at energy $h \nu$, and $M_{k'Q,k}^{e \gamma}$ ($M_{k',kQ}^{e \gamma}$) is the matrix element for absorption of a photon with an electronic transition between states $k$ and $k'$ and concurrent emission (absorption) of a phonon in state $Q$.  In writing Eq.~(\ref{19}) several approximations have been made: (1) the phonon energy is neglected; (2) the photon momentum is neglected; (3) photon emission is omitted, owing to the relatively small excitation levels of interest; (4)  multiphoton absorption is ignored, due to the laser-intensity range of interest; and (5) the distribution of photon energies inherent in a femtosecond pulse is ignored since the width of the distribution is typically $< \,$0.1 eV.  As with Eq.~(\ref{3}), an implied reciprocal lattice vector is present in the Kronecker delta functions $\delta_{\K,\K'+\Q}$ and $\delta_{\K',\K+\Q}$.

In passing, we note spontaneous photon emission by an excited carrier distribution is so weak that it makes negligible contribution to time dependent changes in $f(\epsilon,t)$.  This can be inferred, for example, from photoluminescence measurements on Au by Suemoto \textit{et al}. \cite{Suemoto2019}, where $\sim 10^{18}$ photons/s are absorbed by the sample, but only $\sim 10^5$ photons/s are detected (in any given measurement) as photoluminescent intensity. 

Although Eq.~(\ref{19}) is very similar to the analogous Eq.~(\ref{3}) for electron-phonon scattering, derivation of a $k$-space averaged equation akin to Eq.~(\ref{5}) is not nearly as straightforward.  This is due to two facts.  First, owing to a photon being absorbed in all scattering processes, no pair of matrix elements in Eq.~(\ref{19}) describe processes that are inverses of each other; all four matrix-element terms are thus unique.  Second, unlike the phonon energy $\hbar \omega_Q$, the photon energy $h \nu$ is not often significantly smaller than the Fermi energy $\ep_F$.  However, as we discuss below, two approximations do indeed lead to a simple $k$-space averaged equation.

We now formally proceed as above for electron-phonon scattering.  That is, we multiply Eq.~(\ref{19}) by
$$
\delta(\ep - \ep_k) \, \delta(\ep' - \ep_{k'}) \, \delta(\Omega - \omega_Q) \, \frac{g^{\uparrow}(\ep_k)}{g^{\uparrow}(\ep)}
$$
and integrate on $\ep_k$, $\ep'$, and $\Omega$ [with use of Eq.~(\ref{6b})], which produces (after evaluation of the $\ep'$ integral)
\begin{align}
\label{20}
\bigg( \! \frac{\partial f(\ep)}{\partial t} \! \bigg)_{\!\! e \gamma} = - \frac{\pi}{\hbar} n_{h\nu}(t) & \bigg( g(\ep + h \nu) f(\ep) [ 1 - f(\ep + h \nu) ] \nonumber \\
	& \times \int d\Omega \, F(\Omega) \Big\{  [N(\Omega) + 1 ] \, M_A^2(\ep+h\nu,\Omega;\ep) + N(\Omega) \, M_B^2(\ep+h\nu;\ep,\Omega)  \Big\} \nonumber \\
	& - g(\ep - h \nu) f(\ep - h \nu) [ 1 - f(\ep)] \nonumber \\
	& \times \int d\Omega \, F(\Omega) \Big\{  [N(\Omega) + 1 ] \, M_C^2(\ep,\Omega;\ep-h \nu) + N(\Omega) \, M_D^2(\ep;\ep - h\nu,\Omega)  \Big\}   \bigg).
\end{align}
Here $F(\Omega)$ is the phonon DOS.  The $k$-space averages of the matrix-element terms are given by
\be{\eq}
\label{21a}
M^2_A(\ep',\Omega;\ep)  = \ds{ \sum_{k,k',Q}   \big| M_{k' Q,k}^{e \gamma}  \big|^2  \, \delta^3  \, \delta_{\K,\K' + \Q} } \bigg/\ds{\sum_{k,k',Q}  \delta^3  },
\en{\eq}
\be{\eq}
\label{21b}
M^2_B(\ep';\Omega,\ep)  = \ds{ \sum_{k,k',Q}   \big| M_{k',kQ}^{e \gamma}  \big|^2  \, \delta^3  \, \delta_{\K',\K + \Q} } \bigg/\ds{\sum_{k,k',Q}  \delta^3  },
\en{\eq}
\be{\eq}
\label{21c}
M^2_C(\ep,\Omega;\ep')  = \ds{ \sum_{k,k',Q}   \big| M_{kQ,k'}^{e \gamma}  \big|^2  \, \delta^3  \, \delta_{\K',\K + \Q} } \bigg/\ds{\sum_{k,k',Q}  \delta^3  },
\en{\eq}
and
\be{\eq}
\label{21d}
M^2_D(\ep;\ep',\Omega)  = \ds{ \sum_{k,k',Q}   \big| M_{k,k'Q}^{e \gamma}  \big|^2  \, \delta^3  \, \delta_{\K,\K' + \Q} } \bigg/\ds{\sum_{k,k',Q}  \delta^3  },
\en{\eq}
where for convenience we have defined
\be{\eq}
\label{21e}
\delta^3 = \delta(\ep - \ep_k) \, \delta(\ep' - \ep_{k'}) \, \delta(\Omega - \omega_Q).
\en{\eq}
\end{widetext} 

Quite obviously, what we have before us is rather more complicated than the $k$-space average we obtained for electron-phonon scattering.  We thus naturally ask can we make any further simplifications?  Let's first consider the pair $M^2_A(\ep+h\nu,\Omega;\ep)$ and $M^2_B(\ep+h\nu;\Omega,\ep)$.  These two terms represent averages of transitions from states with energy $\ep$ to states with energy $\ep + h \nu$ with emission and absorption of a phonon with energy $\hbar \Omega \, (\ll h\nu, \ep_F)$, respectively.  It seems reasonable to suppose -- as in the case of electron-phonon scattering -- that these two terms are approximately equal to each other.  For the same reason it is reasonable to suppose $M_C^2(\ep,\Omega;\ep-h \nu) \approx M_D^2(\ep;\ep - h\nu,\Omega)$.  Let's now consider the pair $M^2_A(\ep+h\nu,\Omega;\ep)$ and $M_C^2(\ep,\Omega;\ep-h \nu)$.  These describe averages of the same process, but with the electronic energies shifted by the photon energy $h\nu$.  Because electronic structure does typically change a significant amount on the order of a near-infrared or visible phonon energy, it is not at all clear that $M^2_A(\ep+h\nu,\Omega;\ep) \approx M_C^2(\ep,\Omega;\ep-h \nu)$ is a reasonable assumption.  Nevertheless, we shall proceed with the ansatz that the matrix-element-term averages are indeed independent of $\ep$ and $\ep'$.  This allows us to define a generic averaged matrix-element term $M^2_{e \gamma}(\Omega)$.\footnote{In principle $M^2_{e \gamma}(\Omega)$ could be calculated from any of Eqs. (\ref{21a}) to (\ref{21d}) by assuming the matrix elements are independent of $k$, $k'$, and $Q$. }

Using this ansatz Eq.~(\ref{20}) becomes the rather simple equation
\begin{widetext}
\begin{equation}
\label{22}
\bigg( \! \frac{\partial f(\ep)}{\partial t} \! \bigg)_{\!\! e \gamma} = - K_{e \gamma} \, n_{h \nu}(t) \Big\{ g(\ep + h \nu) f(\ep) [ 1 - f(\ep + h \nu) ] - g(\ep - h \nu) f(\ep - h \nu) [ 1 - f(\ep) ]    \Big\},
\end{equation}
\end{widetext}
where
\begin{equation}
\label{23}
K_{e \gamma} = \frac{\pi}{\hbar} \int d\Omega \, [ 2 N(\Omega) + 1] \, F(\Omega) \, M_{e \gamma}^2(\Omega).
\end{equation}

In order to apply this description of laser excitation to any experimental results, we must first relate the theoretical product $K_{e \gamma} \, n_{h\nu}(t)$ in Eq.~(\ref{22}) to the experimental absorbed intensity $I_a(t)$ [J s$^{-1}$m$^{-2}$] associated with an ultrafast laser pulse.  We do this by considering the change in the density [m$^{-3}$] of excited electrons $n_e$.  Given that $n_e$ can be written as
\be{\eq}
\label{23b}
n_e = \frac{1}{V} \int_{\ep_F}^\infty d\ep \, g(\ep) f(\ep),
\en{\eq}
where $V$ is the volume of the sample, the change in the density of excited electrons (due laser-pulse excitation) can be written as
\be{\eq}
\label{23c}
\frac{d n_e}{dt} = \frac{1}{V} \int_{\ep_F}^\infty d\ep \, g(\ep) \bigg( \! \frac{\partial f(\ep)}{\partial t} \! \bigg)_{\!\! e \gamma}.
\en{\eq}
The connection of this expression to the parameters associated with an ultrafast laser pulse is fairly straightforward.  We first assume we are able to characterize the laser pulse by its absorbed intensity $I_a(t)$ and absorbed fluence $F_a = \int I_a(t) \, dt$ [J m$^{-2}$].  In passing -- and for use in our Au modeling below -- we note for a Gaussian laser pulse the intensity and fluence are related via
\be{\eq}
\label{32b}
I_a(t) =  \frac{2 F_a}{\tau_p} \sqrt{\frac{\ln(2)}{\pi}} e^{-4 \ln(2) (t / \tau_p)^2 },
\en{\eq}
where $\tau_p$ is the duration (FWHM) of the pulse.  We next assume (i) the sample is a thin film with thickness $d_s$, (ii) the light is absorbed uniformly within the depth of the sample, and (iii) each absorbed photon excites an electron from below $\ep_F$ to above $\ep_F$.  Then the rate of electronic excitation is also given by
\be{\eq}
\label{26b}
\frac{dn_e}{dt} = \frac{I_a(t)}{h \nu \, d_s}.
\en{\eq}
If we were to substitute Eq.~(\ref{22}) into Eq.~(\ref{23c}) and equate that relation to Eq.~(\ref{26b}), we would then obtain a relationship between $K_{e \gamma} \, n_{h\nu}(t)$ and $I_a(t)$.  However, that relationship would not be useful, as it involves the time dependent distribution function $f(\ep)$.  However, for our purpose of relating $K_{e \gamma} \, n_{h\nu}(t)$ and $I_a(t)$ it is sufficient to approximate the statistical factors in Eq.~(\ref{22}) by their zero-temperature values.  Doing so and thence utilizing Eqs.~(\ref{22}), (\ref{26b}), and (\ref{23c}) as so described yields the desired relationship
\be{\eq}
\label{28b}
\frac{I_a(t)}{h \nu \, d_s} = \frac{K_{e \gamma} \, n_{h\nu}(t)}{V} \int_{\ep_F}^{\ep_F + h \nu} d\ep \, g(\ep) \, g(\ep - h \nu).
\en{\eq}

This last expression allows us to readily rewrite Eq.~(\ref{22}) in terms of $I_a(t)$.  Solving Eq.~(\ref{28b}) for $K_{e \gamma} \, n_{h\nu}(t)$ we transform Eq.~(\ref{22}) into
\begin{widetext}
\begin{equation}
\label{29b}
\bigg( \! \frac{\partial f(\ep)}{\partial t} \! \bigg)_{\!\! e \gamma} = - \frac{I_a(t)}{h\nu \, d_s} \frac{1}{J(h\nu)} \big\{ g_v(\ep + h \nu) f(\ep) [ 1 - f(\ep + h \nu) ] - g_v(\ep - h \nu) f(\ep - h \nu) [ 1 - f(\ep) ]    \big\},
\end{equation}
where
\begin{equation}
\label{30b}
J(h\nu) = \int_{\ep_F}^{\ep_F + h \nu} d\ep \, g_v(\ep) \, g_v(\ep - h \nu).
\end{equation}
[Recall $g_v(\ep) = g(\ep) /V $ is the DOS per volume.]  If $g_v(\ep)$ is independent of $\ep$ (over the relevant range of $\ep$), then $J(h\nu) = g_v^2 \, h\nu$ and Eq.~(\ref{29b}) thence simplifies to
\begin{equation}
\label{31b}
\bigg( \! \frac{\partial f(\ep)}{\partial t} \! \bigg)_{\!\! e \gamma} = - \frac{I_a(t)}{(h\nu)^2 \, d_s \, g_v} \big\{ f(\ep) [ 1 - f(\ep + h \nu) ] - f(\ep - h \nu) [ 1 - f(\ep) ]    \big\}.
\end{equation}
\end{widetext}
We note this last equation should be quite accurate for any of the noble metals when the photon energy $h \nu$ is small enough that $d$-band excitation can be ignored.  

Somewhat surprisingly, many theoretical treatments of ultrafast electron dynamics in metals do not include $(\partial f / \partial t)_{e \gamma}$ in the BTE equation.  However, when this term is considered, equations very much in line with Eq.~(\ref{22}) are generally used \cite{Sun1994,moore1999,dobryakov1999,Carpene2006,Fatti2000,Kaiser2000,Knorren2000,Rethfeld2002,pietanza2007,Ono2020}, although sometimes a constant DOS $g(\ep)$ is assumed \textit{a priori} \cite{Sun1994,moore1999,dobryakov1999,Carpene2006}.  Several researchers have also included stimulated photon emission in $(\partial f / \partial t)_{e \gamma}$ \cite{Lugovskoy1999,Grua2003}, but (as mentioned above) for the levels of excitation discussed here this process is entirely negligible.

With regard to plasmonic systems, we note that the excitation of plasmons by incident radiation and their subsequent decay is intermediary to the creation of single-particle excitations.  That is, in these systems plasmons are initially excited, but within a few fs they decay into a spectrum of single-electron excitations.  Theoretical studies of plasmon excitation and decay indicate that in many cases this nascent single-electron spectrum is quite similar to that produced by single-electron excitation with no intermediary plasmons\cite{Kornbluth2013,Zhang2014,Bernardi2015,brown2016a,Besteiro2017,Dal2018}.  Our electron-photon scattering integral is thus suitable for application to these systems, as is done in Sec.~\ref{SecRaman}.

\begin{table}[t!]
\footnotesize
\caption{Experimental and DFT-calculated values of the electron-phonon coupling constant $G_{ep}$ [see Eq.~\ref{12b}] for Au. Listed values of $G_{ep}$ are for lowest excitation levels, where applicable.  Experimental technique abbreviations:  TR = transient reflectivity, PTD = photothermal deflection, TSPE = transient surface-plasmon excitation, TT = transient transmission, TED = transient electron diffraction, BiM = transient reflectivity from a bimetallic Pt-Au film, TXRD = transient X-ray diffraction, TSTR= transient spatiotemporal reflectivity. All theoretical studies are density-functional perturbation theory.  Theoretical technique abbreviations refer to the implemented exchange/correlation approximation (when reported). \label{table3b}}
\vspace{0.2cm}
\begin{tabular}{l@{\hspace{0.35cm}}l@{\hspace{0.25cm}}l@{\hspace{0.1cm}}d@{\hspace{0.5cm}}r@{\hspace{0.5cm}}r@{\hspace{0.5cm}}r@{\hspace{0.5cm}}r@{\hspace{0.5cm}}r@{\hspace{0.5cm}}r}

\\

\hline
\hline  
    
Study	 & Technique  	& $G_{ep}$   \\

 &  & $(10^{16} {\, \rm W} {\rm m}^{-3} {\rm K}^{-1})$     \\

\hline

\multicolumn{3}{c} {Experimental}		 \\	

Schoenlein (1987)\cite{Schoenlein1987}  	& TR 	& 2.7$\,$\footnote{Value estimated by Holfield [\onlinecite{Hohlfeld2000}]. \label{holfield}} \\

Brorson (1987)\cite{Brorson1987}		& TR		& 1.9 - 4.1\footnote{Values extracted by Orlande [\onlinecite{Orlande1995}].}	\\

Brorson (1990)\cite{Brorson1990}		& TR		& 2.6 $\pm$ 0.4\textsuperscript{\ref{holfield}}  \\

Elsayed-Ali (1991)\cite{Elsayed-Ali1991}	& TR		& 3.0 - 4.0  \\

Wright (1994)\cite{Wright1994}		& PTD	& 1.6 $\pm$ 0.6  \\

Groeneveld (1995)\cite{Groeneveld1995}	& TSPE	& 3.0 $\pm$ 0.5  \\

Hostetler (1999)\cite{Hostetler1999}	& TR		& 2.9 - 5.0 \\

Hohlfeld (2000)\cite{Hohlfeld2000}		& TR/TT		& 2.0 - 2.3 \\

Smith (2001)\cite{Smith2001}		& TR		& 2.1 $\pm$ 0.2  \\

Ibrahim (2004)\cite{Ibrahim2004}		& TR		& 2.3 \\

Chen (2006)\cite{Chen2006}		& TR		& 2.2 \\

Hopkins (2007)\cite{Hopkins2007} 	& TR		& 2.4 - 2.5 \\

Ligges (2009)\cite{Ligges2009}		& TED	& 2.1 \\

Ma (2010)\cite{Ma2010}			& TR		& 2.07 $\pm$ 0.16\footnote{Average and standard deviation of $G_{ep}$ obtained from data on five samples with different thicknesses.} \\

Hopkins (2011)\cite{Hopkins2011}		& TR		& 2.1 - 2.3 \\

Wang (2012)\cite{Wang2012}		& BiM	& 2.8 \\

Hopkins (2013)\cite{Hopkins2013}		& TR		& 2.3 - 2.5 \\

Choi (2014)\cite{Choi2014}		& BiM	& 2.2 \\

Guo (2014)\cite{Guo2014}		& TR		& 1.5 \\

White (2014)\cite{White2014}		& TXRD	& 2.0 $\pm$ 1.2 \\

Giri (2015)\cite{Giri2015b}		& TR		& 2.3 \\

Naldo (2020)\cite{Naldo2020}		& TR		& 2.4 \\

Sielcken (2020)	\cite{Sielcken2020}		& TR/TT	& 1.97 $\pm$ 0.15 \\

Tomko (2021)\cite{Tomko2021}		& TR		& 2.0 - 2.1 \\

Segovia (2021)\cite{Segovia2021}		& TSTR	& 2.1 $\pm$ 0.4 \\


\\

\multicolumn{3}{c} {Theoretical}		  					 \\

Holst (2014)\cite{Holst2014}				& LDA	& 2.5   \\

Bernardi (2015)\cite{Bernardi2015}	  	& LDA			& 2.2\footnote{Determined from the reported electron-phonon contribution $\tau_{ep}$ to the momentum relaxation time and the relation $G_{ep} = \hbar^2 \langle \Omega^2 \rangle g_v(\ep_F) / (2 \, T_p \tau_{ep})$. \label{bernadi} }  \\

Jain (2016)\cite{Jain2016}		& LDA			& 2.2  \\

Gall (2016)\cite{Gall2016}				& PBE GGA			& 1.9\textsuperscript{\ref{bernadi}} \\

Brown(2016a)\cite{brown2016a}		& PBEsol/DFT+U	 & 2.0\textsuperscript{\ref{bernadi}} \\

Brown (2016b)\cite{Brown2016b}		& PBEsol/DFT+U	& 2.45 \\

Ritzmann (2020)\cite{Ritzmann2020}	&    & 2.25 \\

Li (2022)\cite{Li2022}			& PBE			& 2.39 \\

Eq.~(\ref{12b})			&		& 2.2\footnote{Determined from Eq.~(\ref{12b}); see text for details.} \\

\hline
\hline

\\

\\

\end{tabular}
\bigskip
\end{table}

\begin{table}[t!]
\footnotesize
\caption{Experimental and DFT-calculated values of the electron-electron scattering time $\tau_{ee}(1 \, \rm{eV})$ for Au. Experimental technique abbreviations:  BET = ballistic-electron transport, BEE = ballistic-electron emission, TSPE = transient surface-plasmon excitation, TR2PPE = time-resolved two-photon photoemission.  Theoretical technique abbreviations provide some indication of the particular DFT implementation; see individual references for details. \label{table3c}}
\vspace{0.2cm}
\begin{tabular}{l@{\hspace{0.35cm}}l@{\hspace{0.35cm}}l@{\hspace{0.1cm}}d@{\hspace{0.5cm}}r@{\hspace{0.5cm}}r@{\hspace{0.5cm}}r@{\hspace{0.5cm}}r@{\hspace{0.5cm}}r@{\hspace{0.5cm}}r}

\hline
\hline  
    
Study	 & Technique  	& $\tau_{ee}(1 \, \rm{eV})$   \\

 &  & (fs)     \\

\hline

\multicolumn{3}{c} {Experimental}		 \\	

Sze (1964)\cite{Sze1964}		& BET		& 300\footnote{From Monte-Carlo analysis in Fig.~16 of Ref.~[\onlinecite{Sze1964}].} \\

Sze (1965)\cite{Sze1965}		& BET		& $\ge$29 $\pm$ 4\footnote{Inequality is owing to possible impurity and defect scattering contributions to the scattering time.} \\

Sze (1966)\cite{Sze1966}		& BET		& 92 $\pm$ 36\footnote{Result from our fitting all data ($\delta \ep$ = 0.35 to 1.75 eV) in Fig.~9 of Ref.~[\onlinecite{Sze1966}].} \\

						&			& 38 $\pm$ 20\footnote{Result from our fitting data with $\delta \ep < $ 1.0 eV in Fig.~9 of Ref.~[\onlinecite{Sze1966}].} \\
						
Bauer (1993)\cite{Bauer1993}	& BEE		& 38 $\pm$ 12  \\

Bell (1996)\cite{Bell1996}		& BEE		& 17 $\pm$ 3  \\

Reuter (1999)\cite{Reuter1999}		& BEE		& 23 \\

Reuter (2000)\cite{Reuter2000}		& BEE		& 44 $\pm$ 9 \\

de Pablos (2003)\cite{dePablos2003} 	& BEE		& 67 \\

Groeneveld (1995)\cite{Groeneveld1995}	& TSPE	& 20 $\pm$ $^{20}_7$\footnote{Obtained from reported $K_{ee} =$ 0.1 $\pm$ 0.05 eV$^{-2}$ fs$^{-1}$.}  \\

Aeschlimann (1996)\cite{Aeschlimann1996}		& TR2PPE	& 74\footnote{Scaled from $\delta \ep =$ 1.3 eV data point on 10 nm pc Au/BaF$_2$ film in Fig.~9 of Ref.~[\onlinecite{Aeschlimann1996}].} \\

Cao (1998)\cite{Cao1998}			& TR2PPE		& 89\footnote{Result from 15 nm Au(111)/mica film.}, 84\footnote{Result from 50 nm Au(111)/mica film.} \\

Aeschlimann (2000)\cite{Aeschlimann2000}		& TR2PPE		& 121\footnote{Obtained from an average of data on 10 and 26 nm Au(111)/NaCl films in Fig.~4 of Ref.~[\onlinecite{Aeschlimann2000}].}\\

Bauer (2015)\cite{Bauer2015}		& TR2PPE	& 75\footnote{Obtained by averaging 0.5 eV $<\delta \ep<$ 1.0 eV data from 100 nm pc Au/Ta film; see Fig.~5.9 of Ref.~[\onlinecite{Bauer2015}].} \\

\\


\multicolumn{3}{c} {Theoretical}		  					 \\

Keyling (2000)\cite{Keyling2000}	  	& PW/GW	& 55\footnote{Value is for electron states with $\K$ in the (110) direction.}  \\

Campillo (2000)\cite{Campillo2000}		& PW/GW	& 88\footnote{Derived from the three lowest-energy data points in the inset of Fig.~2 of Ref.~[\onlinecite{Campillo2000}].}  \\

Zhukov (2001)\cite{Zhukov2001}		& LMTO/GW	& 56 $\pm$ 10\footnote{Inferred from two data points closest to $\delta \ep =$ 1.0 eV in Fig.~6 of Ref.~[\onlinecite{Zhukov2001}].  Range estimated from fluctuations in $\tau_{ee}$ near $\delta \ep=$ 1 eV.}  \\

Ladst\"{a}dter (2003)\cite{Ladstadter2003}		& LAPW/GW	& 60 $\pm$ 5\footnote{Obtained from data points with 0.95 eV $\le \delta \ep \le$ 1.05 eV in Fig.~1 of Ref.~[\onlinecite{Ladstadter2003}].} \\

Ladst\"{a}dter (2004)\cite{Ladstadter2004}		& LAPW/GW	& 56 $\pm$ 8\footnote{Obtained from data points with 0.95 eV $\le \delta \ep \le$ 1.05 eV in Fig.~7 of Ref.~[\onlinecite{Ladstadter2004}].} \\

Zhukov (2006)\cite{Zhukov2006}		& GW	& 90\footnote{From Fig.~9 in Ref.~[\onlinecite{Zhukov2006}] \label{zhukov}}  \\

								& GW+T	& 31\textsuperscript{\ref{zhukov}}  \\
								
Brown (2016b)\cite{Brown2016b}		& PW/DFT+U	& 41\footnote{Inferred from reported value of $D_e = \hbar K_{ee}/2 =$ 0.016 eV$^{-1}$.}  \\

\hline
\hline

\\

\end{tabular}
\bigskip
\end{table}

\section{Scattering Strengths in Gold}
\label{SECIII}

Before we proceed with application of the BTE to experimental studies of Au, it behooves us to review prior results related to electron-phonon and electron-electron scattering strengths.  In terms of the BTE theory these strengths are directly characterized by $\lambda \langle \Omega^2 \rangle$ and $K_{ee}$, respectively.  However, different descriptors are often used for these strengths.  As far as electron-phonon scattering is concerned, the electron-phonon coupling constant $G_{ep}$ -- which is related to $\lambda \langle \Omega^2 \rangle$ via Eq.~(\ref{12b}) -- is typically reported.  Similarly, as far as electron-electron scattering is concerned, most studies report values of $\tau_{ee}(\delep)$  -- related to $K_{ee}$ via Eq.~(\ref{29}) -- for some set of $\delep$ values.  Where appropriate, we henceforth employ $G_{ep}$ and $\tau_{ee}$(1$\,$eV) as surrogates for $\lambda \langle \Omega^2 \rangle$ and $K_{ee}$.

Table \ref{table3b} summarizes experimental \cite{Schoenlein1987,Brorson1987,Orlande1995,Brorson1990,Elsayed-Ali1991,Wright1994,Groeneveld1995,Hostetler1999,Hohlfeld2000,Smith2001,Ibrahim2004,Chen2006,Hopkins2007,Ligges2009,Ma2010,Hopkins2011,Wang2012,Hopkins2013,Choi2014,Guo2014,White2014,Giri2015b,Naldo2020,Sielcken2020,Tomko2021,Segovia2021} and theoretical \cite{Holst2014,Bernardi2015,Jain2016,Gall2016,brown2016a,Brown2016b,Ritzmann2020,Li2022} results for $G_{ep}$.  Taken as a whole, the experimentally derived and theoretically calculated values indicate $G_{ep} = 2.2 \times 10^{16}$ W m$^{-3}$ K$^{-1}$ is entirely reasonable.  In fact, this value is obtained using Eq.~(\ref{12b}) with Grimvall's recommendation of $\lambda = 0.17$ \cite{Grimvall1981}, $\hbar^2 \langle \Omega^2 \rangle =  113.6$ eV$^2$ (see Appendix \ref{Appendix B}), and $g_v(\ep_F) = 17.3$ eV$^{-1}$nm$^{-3}$, which comes from the band-structure calculations of Papaconstantopoulos \cite{papaconstantopoulos1986}.   Furthermore, using these values for $\lambda$ and $\hbar^2 \langle \Omega^2 \rangle$, we immediately have that $\lambda \hbar^2 \langle \Omega^2 \rangle =  19.3$ eV$^2$ is equivalent to $G_{ep} = 2.2 \times 10^{16}$ W m$^{-3}$ K$^{-1}$.  We utilize this value for the electron-phonon scattering strength in our BTE calculations below.

As shown in Table \ref{table3c}, electron-electron scattering has also been extensively investigated.  Experimentally, two different ballistic-electron techniques have been utilized.  The earliest work by Sze and coworkers \cite{Sze1964,Sze1965,Sze1966} using ballistic-electron transport (BET) yields $\tau_{ee}$(1$\,$eV) in the range of a few tens to a few hundred fs.  Subsequent ballistic-electron emission (BEE) measurements have been interpreted to give a narrower range between $\sim$ 15 and 65 fs \cite{Bauer1993,Bell1996,Reuter1999,Reuter2000,dePablos2003}.  A third experimental technique -- (low-intensity, single-color) time-resolved two-photon photoemission (TR2PPE) -- yields values within a somewhat higher range than those extracted from BEE,  $\sim$ 75 to 120 fs \cite{Aeschlimann1996,Cao1998,Aeschlimann2000,Bauer2015}.  Lastly, Groeneveld \textit{et al.}~interpret data from their optical surface-plasmon technique to find $\tau_{ee}$(1$\,$eV) = 20 fs, although the reported uncertainty in this value is rather large \cite{Groeneveld1995}.  Theoretically, values between $\sim$ 30 and 90 fs have been calculated \cite{Keyling2000,Campillo2000,Zhukov2001,Ladstadter2003,Ladstadter2004,Zhukov2006,Brown2016b}.  In the following BTE calculations pertinent to fs thermionic-emission measurements \cite{wang1994} we assume $\tau_{ee}$(1$\,$eV) = 75 fs (equivalent to $K_{ee} = 0.027$ eV$^{-2}$fs$^{-1}$), which is the most recent value from TR2PPE data \cite{Bauer2015}.  In modeling data from the other two experimental studies \cite{Fann1992,Huang2014}, we also employ $\tau_{ee}$(1$\,$eV) = 75 fs, but only in our initial analyses.  As shown below, a somewhat smaller value of $\tau_{ee}$(1$\,$eV) is indicated by the data from those studies.

\section{Applications to Experimental Results on Gold}
\label{SectionV}

We now apply our BTE model to previous experiments on Au.  After discussing the applicability of the 2T limit of the model to fs-pulse induced thermionic emission \cite{wang1994}, we model time-resolved photoemission \cite{Fann1992} and ultrafast-pulse Raman scattering \cite{Huang2014} measurements.  For these calculations we employ Eqs.~(\ref{8}), (\ref{17}), and (\ref{29b}) to describe the three relevant BTE scattering integrals.  However, because the Au electronic density of states $g(\ep)$ is nearly constant from the top of the $5d$-band at $\delta \ep \approx - 2.0$ eV to at least $\delep = 6$ eV \cite{papaconstantopoulos1986}, we make the approximation that $g(\ep)$ is constant in these three equations.  We note that with this approximation the relaxation of $f(\ep,t)$ is solely governed by the phonon temperature $T_p$ and the two scattering-strength parameters $\lambda \langle \Omega^2 \rangle$ and $K_{ee}$.

\subsection{Femtosecond thermionic emission}
\label{SecV.B}

\begin{figure*}[t!]
\centerline{\includegraphics[scale=0.65]{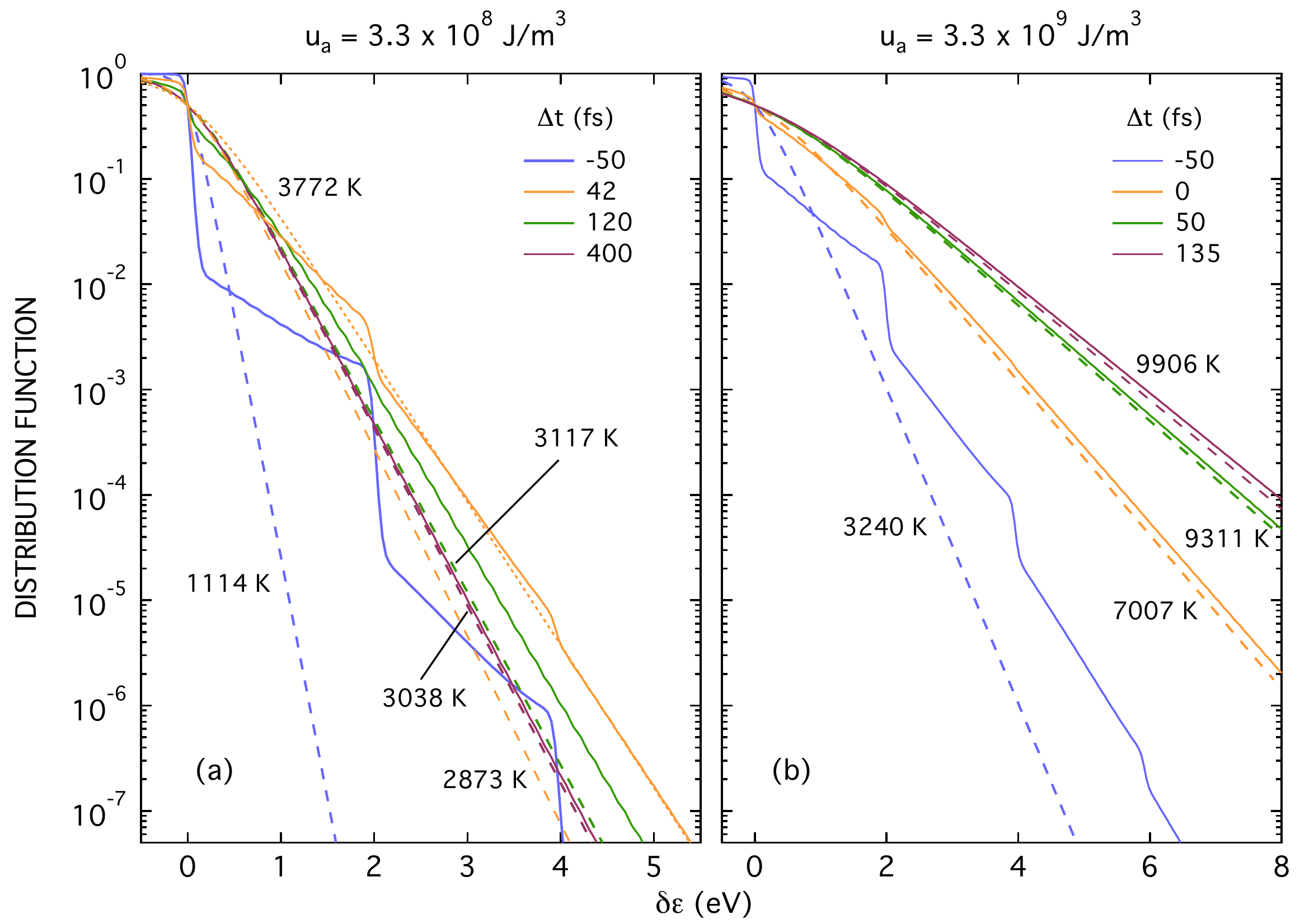}}
\caption{Model distribution functions for Au pertinent to transient thermionic-emission measurements of Wang \textit{et al.}\cite{wang1994}  Solid-curve and dashed-curve distribution functions are calculated using the BTE and 2T models, respectively, and the 2T-model (FD) dashed-line distributions are labeled with the corresponding electron temperature $T_e$.  We note the color of each curve corresponds to a particular time delay, as given by the legend in each panel.  [For example, in panel (a) the solid and dotted green curves correspond to BTE and 2T-model curves at a time delay of 120 fs.]   The dotted curve in panel (a) is a FD distribution at 3772 K.  The time delay $\Delta t = 0$ corresponds to the midpoint of the exciting laser pulse.}
\label{Fig13}
\end{figure*}

Wang \textit{et al}.~performed time-resolved thermionic-emission experiments to investigate carrier dynamics in highly excited Au \cite{wang1994}.  Those experiments utilized $\sim$ 100 fs, 630-nm laser pulses to achieve peak electron temperatures $T_{\rm peak}$ in the range of $\sim$ 3000 to 11,000 K.  Because thermionic emission depends exponentially on $1/T_e$, the value of $T_{\rm peak}$ is the apparent key quantity in these experiments.  By utilizing the 2T model \cite{anisimov1974} in analysis of their data, Wang \textit{et al}.~tacitly assume the carriers can -- at all times -- be described by a thermal distribution.  However, because thermionic emission arises from electrons in the high-energy tail of the distribution, it is quite sensitive to details of the distribution in this region.  Hence, a critical assessment of the thermal-distribution assumption of Wang \textit{et al.} \cite{wang1994} is warranted.  

We have therefore calculated time-dependent distribution functions for Au assuming laser-pulse excitation appropriate to the thermionic-emission experiments.  For these calculations we assume the electron-phonon and electron-electron scattering strengths in Au are characterized by $G_{ep} = 2.2 \times 10^{16}$ W m$^{-3}$ K$^{-1}$ and $\tau_{ee}$(1$\,$eV) = 75 fs, respectively (see the discussion in Sec.~\ref{SECIII}). The results of calculations at absorbed energy densities $u_a = 3.3 \times 10^8$ and $3.3 \times 10^9$ J/m$^3$ are presented in Fig.~\ref{Fig13}.  These two levels were chosen so that the values of $T_{\rm peak}$ are close to the lowest and highest peak temperatures in the thermionic-emission experiments.  More accurate modeling at these excitation levels would incorporate an accurate electronic DOS (owing to thermal excitation of the $5d$ electrons at the highest temperatures in this study), lattice heating, and electron transport.  However, our goal here is to simply compare (very) early-time BTE-model and 2T-model distributions over the relevant range of excitation level.  We do not expect more sophisticated modeling to significantly alter our findings discussed below, and so we proceed with the calculations as described.

Before discussing details of the time development of $f(\ep,t)$, we note at early times $f(\ep,t)$ exhibits a series of steps above $\ep_F$ [$\delep > 0$, see the $-50$ fs (solid blue) curves in Fig.~\ref{Fig13}], with the width of each step being equal to the photon energy $h \nu = 1.97$ eV.  The lowest-energy step is associated with carriers excited by one photon, while the higher energy steps arise from electrons that are sequentially excited by two or more photons \cite{lugovskoy1994,Rethfeld2002}.  

Due to their relative simplicity, we first consider the higher-excitation-level results, displayed in panel (b) of Fig.~\ref{Fig13}.  As is evident, intracarrier thermalization occurs rather quickly:  by the time $\Delta t = 0$ fs the distribution is essentially thermalized and nearly equal to a FD distribution described by the 7007 K temperature calculated using the 2T model (orange curves). This near equivalence continues.  At $\Delta t = 135$ fs (dark red curves) the 2T-model temperature peaks at 9906 K, which is only 2\% less than the equivalent temperature of 10102 K exhibited by the distribution calculated using the BTE.

The results at lower excitation, displayed in  panel (a) of Fig.~\ref{Fig13}, are more intriguing.  As is evident by comparison with the results in panel (b), intracarrier thermalization at this lower level of excitation takes much longer.  However, the key result regarding thermionic emission is evident in the three (orange) curves related to $\Delta t = 42$ fs.  At this time delay the BTE and 2T-model calculated ($T_e =$ 2873 K) distributions are still quite divergent.  However, the highest energy part of the BTE calculated distribution is essentially thermalized, being well described by $T_e = 3772$ K, as illustrated by comparison with the FD distribution (dotted curve) at this temperature.  Importantly, this (maximum) effective temperature is significantly larger than 2T-model peak temperature of 3117 K, which occurs at 120 fs (dashed green curve).  We note that by $\Delta t = 400$ fs the carriers are essentially thermalized at all energies and can be characterized by $T_e$ very close to the 2T-model temperature of 3038 K (dark red curves).

The ratio 1.21 of the effective peak temperature of 3772 K to the 2T-model peak temperature of 3117 K explains an incongruence in experimental and 2T-model calculated temperatures reported by Wang \etal \cite{wang1994} As shown in Fig.~2 of that paper, the experimental and theoretical values of $T_{\rm peak}$ agree quite well at the three highest excitation levels.  However, at the lowest excitation level a significant difference exists, with the 2T-model value of $T_{\rm peak}$ appreciably below the experimental value.  In that experiment $T_{\rm peak}$ is measured as a function of the time displacement between two equal-energy excitation laser pulses.  For zero time displacement the lowest-excitation experimental and 2T-model values of $T_{\rm peak}$ are 3410 and 2940 K, respectively.  The ratio of these two temperatures is 1.16, quite close to the ratio 1.21 noted above.

Peak carrier temperatures similar to those in the Wang \etal~experiment have been generated in a number of other investigations that utilize fs laser pulses, including studies of transient-grating generation and decay \cite{Hibara1999}, x-ray production \cite{Egbert2002}, electric-field-enhanced thermionic emission \cite{Wang2015}, ultrafast-laser damage thresholds, \cite{Chen2010} and laser ablation \cite{Li2015}.  In each of these cited studies the 2T model is used to interpret the results.  Given our calculations for Au, we suggest critical assessment of the 2T model may also be warranted before being applied to experiments such as these, especially when the details of the distribution of the highest energy electrons is important.

\subsection{Transient photoemission measurements}
\label{SecVIB}

\begin{figure*}[t]
\centerline{\includegraphics[scale=0.58]{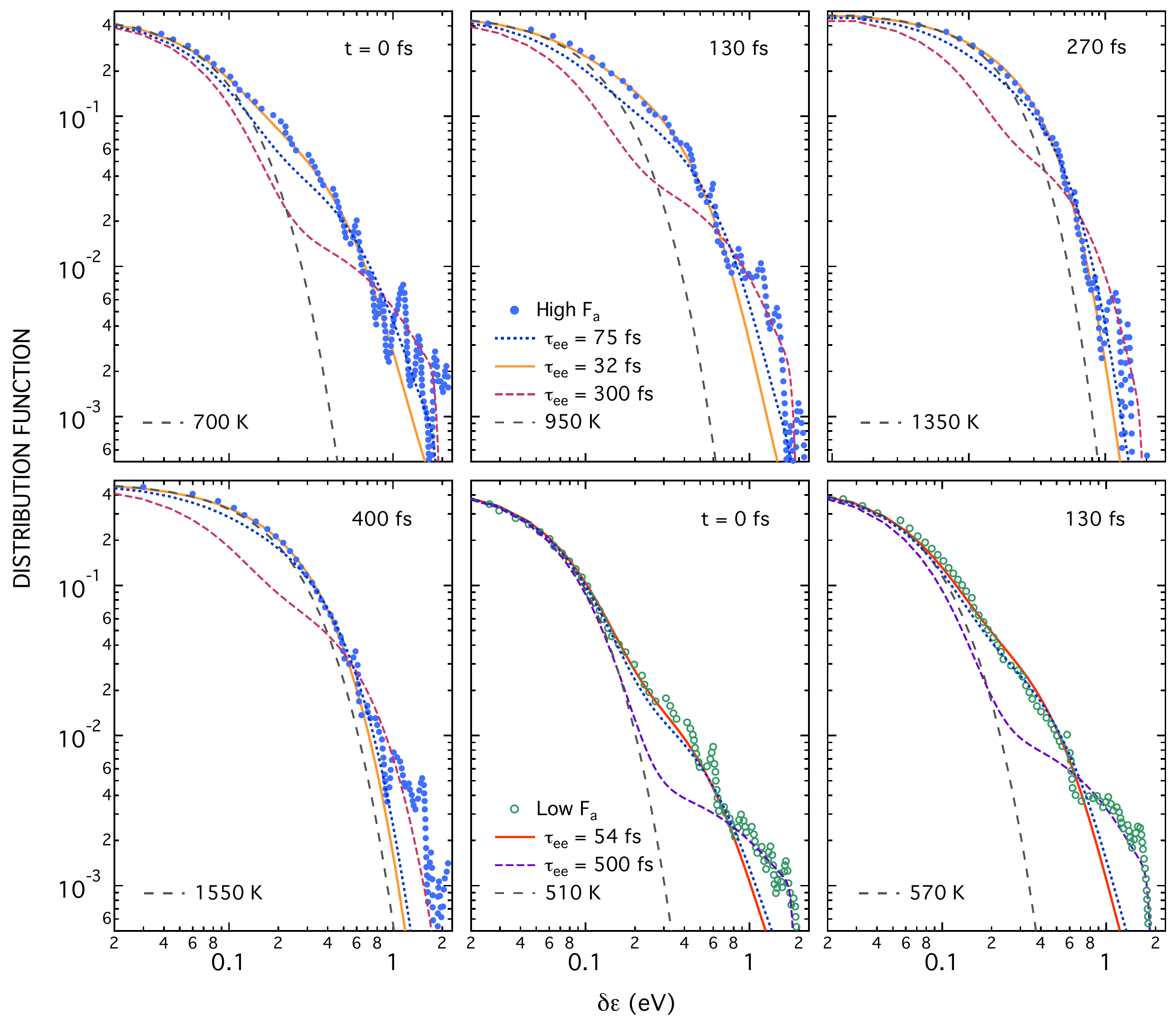}}
\caption{Probe-pulse-convoluted distribution functions for 1.84 eV transient excitation of Au.  Circles are experimental results from Fann \textit{et al.}\cite{Fann1992}  Solid (blue) circles in the first four panels are at high-fluence ($F_a = 300 \, \pm \, 90 \, \mu$J) and open (green) circles are at low-fluence ($120 \, \pm \, 40 \, \mu$J).   Dotted, solid, and short-dash curves are BTE model calculations with $\tau_{ee}$(1$\,$eV) as indicated.  Long-dash curves are FD distributions at the indicated temperatures.  These FD curves match the solid curves at the lowest energies.  For each data set the time interval between probe and pump pulses is noted.}
\label{Fig14}
\end{figure*}

Perhaps the most iconic experiments associated with nonthermal carriers are the time-resolved, two-color photoemission measurements of Fann \etal \cite{Fann1992}  In these experiments a 30 nm Au film is excited by a 1.84 eV, 180 fs laser pulse.  A variable-delay 5.52 eV, 270 fs probe pulse is then used to photoemit electrons from the excited distribution.  From these measurements Fann \textit{et al.}~construct two sequences of five (probe-pulse convoluted) carrier distributions each, one set with absorbed laser fluence $F_a = 120 \pm 40$ $\mu$J cm$^{-2}$ and another at $F_a = 300 \pm 90$ $\mu$J cm$^{-2}$.  Both sets exhibit clear evidence for the nonthermal nature of the carriers.  While subsequent two-color photoemission studies have revealed similar nonthermal distributions [on supported Ag nanoparticles \cite{Merschdorf2002,Pfeiffer2004}, Ru(001) \cite{Lisowski2004}, superconducting Bi$_2$Sr$_2$CaCu$_2$O$_{8+\delta}$ \cite{Perfetti2007}, graphite \cite{Rohde2018,Na2020}, and Pb/Si(111) \cite{Kratzer2022}], none of these later studies have produced as extensive an array of distributions.

No doubt owing to the richness of the Fann \etal~data, there have been several attempts to model their experimental distributions:  in addition to a model put forth by Fann \etal~themselves, Bejan and Ra{\c{s}}eev \cite{Bejan1997} and Lugovskoy and Bray \cite{Lugovskoy1999} also calculate sets of theoretical distributions appropriate to the experiment.  Unfortunately, all three models lack key features.  First, none of these models utilize a full-integral description of electron-electron scattering [such as that expressed by Eq.~(\ref{17})].  Rather, each model simply treats this interaction using single-carrier scattering in the relaxation-time approximation (RTA), with the relaxation time given by the equivalent of Eq.~(\ref{29}).  Second, in none of these models does electron-phonon scattering directly alter the entire distribution function.  Fann \etal \cite{Fann1992} and Lugovskoy and Bray \cite{Lugovskoy1999} indirectly include this interaction via a calculation that involves the 2T-model in some way, while Bejan and Ra{\c{s}}eev \cite{Bejan1997} ignore this interaction altogether.  Third, none of the calculations account for the experimental reality that a 270 fs probe pulse is used to sample the pump-pulse excited distributions.  The models exhibit variable success in matching the experimental distributions;  that of Lugovskoy and Bray is the most accurate.  However, one must question the values of any physical parameters used in any of the models, given their simplicity.  For example, the model of Lugovskoy and Bray utilizes an electron-phonon coupling constant $G_{ep}$ equal to $2.7 \times 10^{17}$ W m$^{-3}$ K$^{-1}$, a value about 10 times larger than that indicated by most experiments and theory (see Sec.~\ref{SECIII} above).  Indeed, using our model we cannot accurately fit the data using this value of $G_{ep}$.   

So what can we learn from the experimental distributions of Fann \textit{et al.}~in the context of a more complete description of carrier dynamics?  Here we attempt to answer this question by carefully comparing those distributions with calculations using our model.  To make these comparisons we have four adjustable parameters:  the absorbed laser fluence $F_a$, the phonon temperature $T_p$, and the electron-phonon and electron-electron scattering strengths, which we continue to characterize using $G_{ep}$ and $\tau_{ee}$(1$\,$eV), respectively.

The results of our analysis are summarized in Fig.~\ref{Fig14}.  Four panels exhibit experimental data (and our corresponding calculations) obtained at $F_a = 300 \pm 90$ $\mu$J cm$^{-2}$, and two panels show data (and calculations) for $F_a = 120 \pm 40$ $\mu$J cm$^{-2}$.  We note our theoretical curves are not simply $f(\delta\ep,t)$ at a particular time delay (as is the case with the three prior analyses \cite{Fann1992,Bejan1997,Lugovskoy1999}), but are rather the convolution of $f(\delta\ep,t)$ with an area-normalized Gaussian pulse of duration 270 fs (equal to the probe-pulse duration), which corresponds to the actual experimental measurement.

Before discussing our analysis we point out all experimental data displayed in Fig.~\ref{Fig14} are indeed nonthermal in nature.  To see this we include in each panel a thermal distribution (with temperatures indicated) as a long-dash curve.  These curves are chosen to match the solid curves from our analysis at energies $\delta \ep \lesssim 0.1$ eV.  In all cases the experimental data lie significantly above these thermal curves at energies $\delta \ep \gtrsim 0.2$ eV.

With regard to our analysis, the first notable observation is the experimental data cannot be theoretically matched over the whole range of $\delta \ep$.  An accurate fitting of the data can be obtained over the range $0 < \delta \ep \lesssim 0.8$ eV (note the solid curves in Fig.~\ref{Fig14}), but such fitting results in the theory being consistently lower than the experimental data at $\delta \ep \gtrsim 0.8$ eV.  This fact was also discerned by Lugovskoy and Bray \cite{Lugovskoy1999}.  Conversely, accurately fitting the higher energy region (note the short-dashed curves) produces theoretical curves consistently lower than the data in the lower energy region.  These observations suggest the presence some systematic error in the experiment. As discussed in more detail below, we believe data in the $0 < \delta \ep \lesssim 0.8$ eV range to be the more reliable.

A second inconsistency occurs among the distribution curves obtained at each fluence.  At the higher fluence we find the four distributions shown in Fig.~\ref{Fig14} are well described by the same set of parameters.  However a fifth experimental distribution (obtained at the probe-pulse delay $t = 670$ fs, not reproduced in the figure), is then significantly below the theory at all energies (see the Supplemental Material).  Similarly, at lower fluence the two earliest time distributions can be well described by the same parameters, but then the three later-time distributions ($t =$ 400, 670, and 1300 fs, also not reproduced here) all trend lower than the theoretical curves (again, see the Supplemental Material).  We therefore focus our attention on the six experimental distributions shown in Fig.~\ref{Fig14}.

Let's first consider the nonintrinsic parameters $F_a$ and $T_p$.  Our analysis extracts values for $F_a$ that are completely consistent with experiment:  we find theoretical curves that best match the data have low and high fluences of $F_a = 103$ and $325$ $\mu$J cm$^{-2}$, respectively.  Conversely, the fitted values of the phonon temperature $T_p$ are a bit puzzling:  in order to have the lowest-energy part ($\delta \ep \lesssim 0.1$ eV) of all distributions be reasonably well fit, values of $T_p$ significantly larger than room temperature (RT) are required.  Our best descriptions of the data require $T_p = 485$ and 540 K for the low-fluence and high-fluence data sets, respectively.  In a report on a preliminary experiment Fann \textit{et al.}~note their photoemission data imply $T_p = 380 \pm 30$ K, also above RT \cite{Fann1992A}.  They attribute this increased temperature to a systematic error associated with the 30 meV energy resolution of their electron spectrometer.  However, we find this broadening can only account for an apparent 14 K increase (above RT) in $T_p$.  Two possible explanations come to mind.  First, a heating source within the ultrahigh vacuum chamber (such as an ionization gauge) might have produced an increase in the base temperature of the Au sample.  Second, $T_p$ being higher for the higher fluence data suggests an accumulation heating effect from the laser pulses might be at least partially responsible.

We now turn our attention to the first of the two intrinsic parameters, the electron-phonon scattering strength.  Because the experimental data of Fann \textit{et al.} \cite{Fann1992} are obtained at rather early times, the data are much less sensitive to this parameter than $\tau_{ee}$.  For example, doubling or halving any reasonable value of $G_{ep}$ is easily compensated in our fitting by a slight increase or decrease in $F_a$.  Owing to this insensitivity, in our analysis we simply set  $G_{ep}$ to the value of $2.2 \times 10^{16}$ Wm$^{-3}$K$^{-1}$.

\begin{figure}[t!]
\centerline{\includegraphics[scale=0.46]{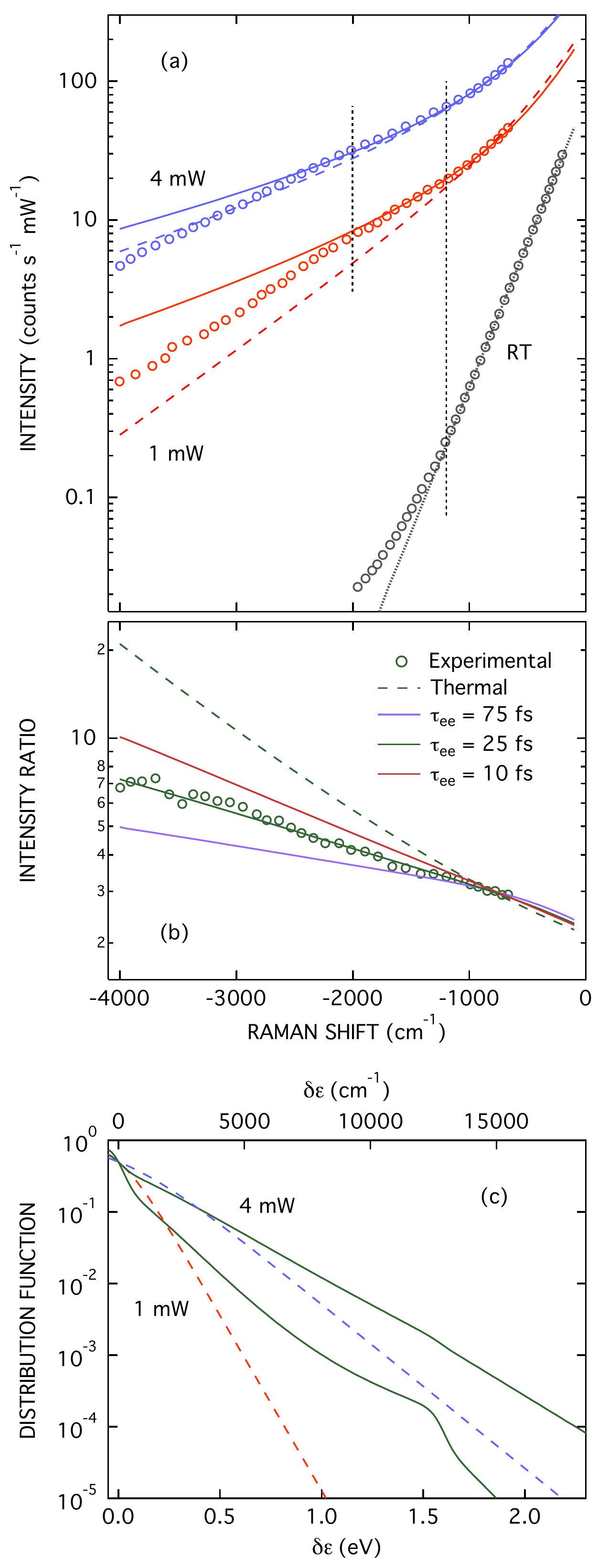}}
\caption{Anti-Stokes ($\omega \! < \! 0$) Raman-scattering from Au nanorods using 785 nm laser light.  (a) Open circles are experimental RS data from Huang \textit{et al.} \cite{Huang2014}  RT data were obtained with cw-laser excitation; the 1 and 4 mW (average-power) data were obtained using 450 fs laser pulses.  Dashed and solid curves are calculations using thermal (FD) and BTE derived distributions, respectively; see text for details.  The thermal curves are for $T_e =$ 1030 K (1 mW) and 2200 K (4 mW), while the BTE theory curves are calculated for $\tau_{ee}$(1$\,$eV) = 75 fs.  The dotted line through the RT data is a simple exponential-decay function.    (b) Intensity ratios of the 4 mW to 1 mW experimental data and calculations.  Calculations using the BTE model for three different values of $\tau_{ee}$(1$\,$eV) are shown.  (c) Dashed and solid curves are FD and BTE calculated distribution functions, respectively.  The FD distributions are for $T_e =$ 1030 K (1 mW) and 2200 K (4 mW).   The BTE distribution functions are laser-pulse-intensity weighted and are calculated for $\tau_{ee}$(1$\,$eV) = 25 fs, which corresponds to the best fit of the experimental data shown in (b). }
\label{Fig15}
\end{figure}

So what do the data of Fann \textit{et al.}~tell us about the electron-electron interaction?  In Fig.~\ref{Fig14} we present calculations using several values of $\tau_{ee}$(1$\,$eV).  We first note $\tau_{ee} =$ 75 fs (the value from the most recent TR2PPE measurements, see Sec.~\ref{SECIII} above) produces curves that do not quite match experiment.  For $\delta \ep \lesssim 0.8$ eV the calculated distributions are slightly less than the experimental distributions, most notably for the high-fluence spectra.  Additionally, for $\delta \ep \gtrsim 0.8$ eV the calculated curves are a poor match, especially at low fluence.  For $\delta \ep \lesssim 0.8$ eV better agreement is obtained with values of $\tau_{ee}$(1$\,$eV) somewhat smaller that 75 fs:  54 and 32 fs at low and high fluence, respectively.  That these two values are not more similar likely reflects limitations of the experimental data.  However, the two values are within the range of prior experimental and theoretical values, as summarized in Sec.~\ref{SECIII}.  Given this agreement, we feel the $\delta \ep \lesssim 0.8$ eV data of Fann \textit{et al.}~are more reliable than those above 0.8 eV.

This point is emphasized by fits that match the experiment in the higher energy range.  As also shown in the figure, a good match to the experiment above $\delta \ep \approx 0.8$ eV is obtained with $\tau_{ee}$(1$\,$eV) = 500 fs at low fluence and 300 fs at high fluence.  These fits, however, are severely below all  experimental curves in the lower energy region.  Because these two values of $\tau_{ee}$(1$\,$eV) are $\sim$ 4 to 10 times most prior results, we conjecture the Fann \textit{et al.}~data in this energy region are plagued by some systematic error.  We offer no potential explanation for such error, but we note similar transient photoemission measurements of Ag nanoparticles require the subtraction of a three-photon background in order to extract the nonthermal carrier distribution \cite{Merschdorf2002,Pfeiffer2004}.  Measurements at a negative time delay might have been helpful in distinguishing any potential background in the Fann \textit{et al.}~data.

\subsection{Subpicosecond Raman scattering}
\label{SecRaman}

We now turn our attention to ultrafast Raman scattering (RS):  using 1.58 eV (785 nm), 450 fs laser pulses Huang \etal~have measured spectra from Au nanorods (AuNRs) at several different  excitation levels \cite{Huang2014}.  Here we show these spectra (i) arise from nonthermal carrier distributions and (ii) are sensitive to the carrier-carrier interaction.  As with the transient photoemission measurements of Fann \etal \cite{Fann1992}, we are able to characterize the data with a scattering time $\tau_{ee}$(1$\,$eV).

The particular data we consider are shown in Fig.~\ref{Fig15}.  Panel (a) plots ultrafast anti-Stokes (AS) spectra obtained at average powers of 1 and 4 mW.  Also shown in this panel are data taken with continuous-wave (cw) radiation; these data are standard RT RS spectra from the same sample.\footnote{The femtosecond-pulse and RT spectra are respectively from Figs.~S5 and 2 of Huang \etal \cite{Huang2014}  Ultrafast data are also displayed in Fig.~2, but the frequency range is much more limited, as it only extends to $-2000$ cm$^{-1}$. We have therefore analyzed the Fig.~S5 ultrafast data.  The data in Figs.~2 and S5 were nominally obtained under the same conditions; however, the overall count rate in Fig.~S5 is lower by a factor of $\sim$ 1.65.  Therefore, in order to place the Fig.~S5 ultrafast data on the same scale as the Fig.~2 RT data, we have multiplied the Fig.~S5 spectra by 1.65. }  We note in the digitization process the RS data are effectively low-pass filtered; hence, counting noise in the original data is not manifest in the experimental curves in Fig.~\ref{Fig15}(a).

In order to model these data we develop a mathematical model of RS appropriate to a time-dependent nonthermal distribution:  we begin with an expression applicable to a thermal distribution, and thence extend that expression to encompass dynamic carriers.  For thermal carriers the RS intensity $I_{\rm RS}^{th}$ at Raman-shift frequency $\omega = \omega_i - \omega_s$ can be generically expressed as\cite{Ipatova1981,Abramsohn1982,Kostur1992,Ponosov2012}
\be{\eq}
\label{92}
\frac{I_{\rm RS}^{th}(\omega,\omega_i,T_e)}{I_i} = F_{\rm RS} \, \sigma_{\rm RS}(\omega,\omega_i) \, \frac{\hbar \omega}{1 - e^{-\hbar \omega / k_B T_e}} .
\en{\eq}
Here $\omega_i$ and $\omega_s$ are the incident and scattering radiation frequencies, respectively, and $\sigma_{\rm RS}(\omega,\omega_i)$ is a material response function that depends not only upon $\omega$ and $\omega_i$ but also the polarizations of the incident and scattered electromagnetic fields.\footnote{Although not universal, the response function $\sigma_{\rm RS}$ can often be approximated by the real part of the Drude conductivity $\tilde{\sigma}(\omega) = \sigma_0 / (1 - i \omega \tau)$. Here $\sigma_0$ is the dc conductivity, and $\tau$ is the momentum relaxation time of the carriers.  See, for example, Refs. [\onlinecite{Kostur1992}] and [\onlinecite{Ponosov2012}]. }  The quantity $I_i$ is the incident light intensity, and the proportionality constant $F_{\rm RS}$ subsumes experimental quantities such as scattering geometry and light transmission at any relevant interfaces.  We now invoke an expression for electronic RS introduced by Jain \etal, \cite{Jain1976}
\be{\eq}
\label{93}
I_{\rm RS}(\omega) \propto \int_{-\infty}^\infty d\ep \, g(\ep) f(\ep) g(\ep + \hbar \omega) [1 - f(\ep + \hbar \omega)].
\en{\eq}
This expression is similar to energy-integral equations above [such as Eqs.~(\ref{8}) and (\ref{17})] in that it derives from assuming $k$-independent matrix elements for the relevant scattering process. If the carriers are thermalized [so that $f(\ep) = f_{\scriptscriptstyle \!F\!D}(\ep,T_e)$] and $g(\ep)$ is constant, then Eq.~(\ref{93}) becomes
\begin{widetext}
\begin{equation}
\label{94}
I_{\rm RS}(\omega)   \propto \int_{-\infty}^\infty d\ep \, \frac{1}{e^{ (\ep - \ep_F) / (k_B T_e)} +1} \bigg[1 - \frac{1}{e^{ (\ep + \hbar \omega - \ep_F )/ (k_B T_e)} +1}\bigg] = \frac{\hbar \omega}{1 - e^{-\hbar \omega / (k_B T_e)}}.
\end{equation}
The expression on right side of this equation appears as a factor in Eq.~(\ref{92}); we are thus motivated to make the ansatz
\begin{equation}
\label{95}
\frac{I_{\rm RS}(\omega,\omega_i,[f]) }{I_i} = \frac{F_{\rm RS}  \, \sigma_{\rm RS}(\omega,\omega_i)}{g^2(\ep_F) } \, \int_{-\infty}^\infty d\ep \, g(\ep) f(\ep) g(\ep + \hbar \omega) [1 - f(\ep + \hbar \omega)]
\end{equation}
for any carrier distribution, thermalized or not.\footnote{Here we use the notation $[f]$, as $I_{RS}$ is a functional of the distribution function $f(\ep)$.}  If it so happens the DOS $g(\ep)$ can be approximated as constant, then Eq.~(\ref{95}) readily reduces to
\begin{equation}
\label{96}
\frac{I_{\rm RS}(\omega,\omega_i,[f])}{I_i} = F_{\rm RS} \, \sigma_{\rm RS}(\omega,\omega_i) \, \int_{-\infty}^\infty d\ep \, f(\ep)  [1 - f(\ep + \hbar \omega)].
\end{equation}
In the experiment of Huang \etal~each incident laser pulse not only creates the evolving carrier distribution, but is also inelastically scattered by that distribution.  To account for these dynamics this last equation naturally extends to
\begin{equation}
\label{97}
\frac{I_{\rm RS}(\omega,\omega_i,[f])}{I_i} = F_{\rm RS} \, \sigma_{\rm RS}(\omega,\omega_i) \, \int_{-\infty}^\infty dt \, I_n(t) \! \int_{-\infty}^\infty d\ep \, f(\ep,t)  [1 - f(\ep + \hbar \omega,t)],
\end{equation}
where $I_n(t)$ is a unit-area function with the shape of the laser-pulse intensity.  We shall use this relation for $I_{RS}(\omega,\omega_i,[f]) / I_i$ in our calculations for AuNRs.  Unfortunately, we are in no position to calculate the product $F_{RS} \, \sigma_{RS}(\omega,\omega_i)$.  To circumvent this issue we analyze ratios of RS intensities, thereby eliminating the need to have any knowledge of $F_{RS} \, \sigma_{RS}(\omega,\omega_i)$.  For example, in the first part of our ensuing analysis we consider RS spectra at two different sample temperatures $T_1$ and $T_2$.  From Eq.~(\ref{92}) we immediately have
\begin{equation}
\label{98}
\frac{I_{\rm RS}^{th}(\omega,T_2) / I_i}{I_{\rm RS}^{th}(\omega,T_1) / I_i} = \frac{1 - e^{-\hbar \omega / k_B T_1}}{1 - e^{-\hbar \omega / k_B T_2}}.
\end{equation}
We note this equation has been previously used to good effect:  by taking the ratio between AS RS spectra from Au nanodots when heated ($T_2$) and unheated ($T_1 = 298$ K), Xie and Cahill demonstrate Eq.~(\ref{98}) can be used to accurately ($\pm2$\%) determine the temperature of the heated nanodots \cite{Xie2016}.  As we show below, other RS-spectra ratios also prove useful.

We first demonstrate the 450-fs-pulse RS spectra (1 and 4 mW) in Fig.~\ref{Fig15} are consistent with a thermal distribution in the frequency range $|\nubar| < 1200$ cm$^{-1}$ [for reference, $\nubar = \omega/(2 \pi c)$].  To do this we utilize the fact that within this wavenumber range the RT RS spectra are phenomenologically well described by a single exponential function,
\begin{equation}
\label{99}
\frac{I_{\rm RS}^{th}(\omega,300\,{\rm K})}{I_i} = 75 \, e^{\hbar \omega / (k_B 300 {\rm K})} \, \rm{counts \,\, s^{-1} \,\, mW^{-1}};
\end{equation}
this equation is plotted as the dotted line in part (a) of Fig.~\ref{Fig15}.  Using this expression in Eq.~(\ref{98}) for the reference intensity $I_{\rm RS}^{th}(\omega,T_1) / I_i$ ($T_1 = 300$ K), we have calculated $I_{\rm RS}^{th}(\omega,T_2) / I_i$ for $T_2 = 1030$ and 2200 K, and plotted the results as the dashed curves in Fig.~\ref{Fig15}(a).  Quite obviously, these curves line up with the experimental data for $|\nubar| < 1200$ cm$^{-1}$.  Satisfyingly, Huang \etal~independently extract the same effective temperatures (to within 5 K) for 1 mW and 4 mW excitation. \{They do this by assessing scattered intensity at $-500$ cm$^{-1}$ (see Fig.~4B of Ref.~[\onlinecite{Huang2014}])\}.  For $|\nubar| > 1200$ cm$^{-1}$ a divergence exists between the experimental data and the thermal-distribution curves.  This difference is partially is due to the deviation of experimental RT RS spectra from Eq.~(\ref{99}) at higher frequencies,\footnote{Clearly the RT RS spectra are not described by the exponential decay of Eq.~(\ref{99}) for $\nubar > 1200$ cm$^{-1}$.  However, we suspect the data in this region are influenced by a nonzero background.  Because of this suspicion and because there are no RT data for $\nubar > 2000$ cm$^{-1}$, we simply use Eq.~(\ref{99}) to model the RT scattering at all frequencies. } but it is also due to the nonthermal nature of the data.

We now model RS spectra from a dynamic distribution excited by a 1.58 eV, 450 fs laser pulse.  For the first part of this analysis we use the ratio of the intensity given by Eq.~(\ref{97}) [with $I_n(t)$ being a 450 fs Gaussian] to that of a thermal distribution, given by Eq.~(\ref{92}).  Using these two equations this ratio is readily expressed as
\begin{equation}
\label{100}
\frac{I_{\rm RS}(\omega,[f_2]) / I_i}{I_{\rm RS}^{th}(\omega,T_1) / I_i} =  \frac{\int_{-\infty}^\infty dt \, I_n(t) \! \int_{-\infty}^\infty d\ep \, f_2(\ep,t)  [1 - f_2(\ep + \hbar \omega,t)]}{\hbar \omega / (1 - e^{-\hbar \omega / k_B T_1})}.
\end{equation}
\end{widetext}
The RT RS data in Fig.~\ref{Fig15}(a) deviate substantially from the exponential description of Eq.~(\ref{99}) for $|\nubar| > 1200$ cm$^{-1}$.  However, because RT data do not exist for $|\nubar| > 2000$ cm$^{-1}$, we continue to use Eq.~(\ref{99}) to describe $I_{\rm RS}^{th}(\omega,T_1) / I_i$ ($T_1 = 300$ K), now in Eq.~(\ref{100}) at all frequencies. 

We now discuss how the Raman-scattering signals depend upon the electron-phonon and electron-electron interaction strengths.  In conjunction with details of the excitation process, these interaction strengths again control the time-evolution of the distribution function $f(\ep,t)$.  The electronic Raman scattering signal itself is a time average of $f(\ep,t)$, weighted by the laser-pulse intensity [see Eq.~({\ref{97})].  Similar to the two-photon photoemission data analyzed in Sec.~\ref{SecVIB}, the ultrafast Raman-scattering data are quite insensitive to the strength of the electron-phonon interaction.  For example, doubling $G_{ep}$ from the value of $2.2 \times 10^{16}$ W m$^{-3}$ K$^{-1}$ established above to $4.4 \times 10^{16}$ W m$^{-3}$ K$^{-1}$ is readily compensated by increasing the absorbed energy density by only 8\%.  Therefore, in the ensuing analysis we again fix $G_{ep}$ at $2.2 \times 10^{16}$ W m$^{-3}$ K$^{-1}$.  Spectra obtained from $f_2(\ep,t)$ calculated with the nominal value $\tau_{ee}$(1$\,$eV) = 75 fs and energy densities $u_d = 1.07$ and 4.40 $\times 10^8$ J m$^{-3}$ (chosen so that the calculations align with the experimental data in the the $|\nubar| < 1200$ cm$^{-1}$ region) are shown in Fig.~\ref{Fig15}(a) as the solid curves.  It is encouraging these nonthermal-distribution spectra match the experimental data better than the FD-distribution spectra in the region up to $|\nubar| = 2000$ cm$^{-1}$.  However, at higher values of $|\nubar|$ deviations continue to exist.  This issue is due at least in part to the lack of a RT RS reference spectrum in this frequency range.

To overcome this limitation we analyze the ratio of the ultrafast 4 mW to 1 mW spectra.  The ratio of the experimental data of Huang \etal~is displayed as the open circles in part (b) of Fig.~\ref{Fig15}.  Shown also in part (b) as the dashed curve is the ratio of the two thermal-distribution calculations in part (a).  The increasing deviation between the experimental and thermal-distribution ratios with increasing $|\nubar|$ clearly indicates the nonthermal-distribution nature of the data.  Ratios of RS intensities for calculations of dynamic distributions are immediately obtained from Eq.~(\ref{97}) and given by
\begin{widetext}
\begin{equation}
\label{101}
\frac{I_{\rm RS}(\omega,[f_2]) / I_i}{I_{\rm RS}(\omega,[f_1]) / I_i} =  \frac{\int_{-\infty}^\infty dt \, I_n(t) \! \int_{-\infty}^\infty d\ep \, f_2(\ep,t)  [1 - f_2(\ep + \hbar \omega,t)]}{\int_{-\infty}^\infty dt \, I_n(t) \! \int_{-\infty}^\infty d\ep \, f_1(\ep,t)  [1 - f_1(\ep + \hbar \omega,t)]}.
\end{equation}
\end{widetext}
The ratio for the two (solid) curves in part (a) -- shown as the solid violet curve in (b) -- is not a bad match to experiment for $|\nubar| \lesssim 1700$ cm$^{-1}$, but at higher frequencies the experimental data are underestimated.  We have thus redone the calculations for several other values of $\tau_{ee}$(1$\,$eV).  As also shown in (b) the solid green curve for $\tau_{ee}$(1$\,$eV) = 25 fs matches the experimental ratio quite well over the whole range of frequencies.  In addition to that for 75 fs, the solid red curve calculated with $\tau_{ee}$(1$\,$eV) = 10 fs gives some indication of the sensitivity of the experimental data to this parameter.

Interestingly, the best-fit value of 25 fs for $\tau_{ee}$(1$\,$eV) is not that far from our best value of 32 fs extracted from the high-fluence photoemission data of Fann \etal \cite{Fann1992} These two results suggest values (75 to 120 fs) extracted from TR2PPE measurements might systematically err on the high side.  However before any such conclusions can be made, further theoretical justification for the use of Eq.~(\ref{97}) or development of a potentially more accurate expression is certainly warranted. Importantly, though, our analysis demonstrates ultrafast Raman scattering is quite sensitive to the early-time dynamics of laser-excited carriers.

This last point is emphasized in panel (c) of Fig.~\ref{Fig15}.  Here we plot laser-pulse weighted distribution functions $\int dt \, I_n(t) f(\ep,t)$ (solid curves) calculated with our BTE model.  These two curves are calculated for $\tau_{ee} =$ 25 fs [and so are associated with the best-fit curve in panel (b)].  As can be seen in comparison with FD distribution functions (dashed curves) calculated at the two temperatures associated with 1 and 4 mW data, the BTE calculated distributions are -- effectively -- significantly hotter.



\section{Summary}

In this paper we accomplish two key objectives regarding application of the Boltzmann transport equation (BTE) to ultrafast carrier dynamics in laser excited metals.  Here we summarize these accomplishments.

(i) In the spirit of the random-$k$ approximation \cite{berglund1964,kane1967} we present parallel derivations of energy dependent electron-phonon, electron-electron, and electron-photon scattering integrals BTE [Eqs.~(\ref{8}), (\ref{17}), and (\ref{22}), respectively].  Each integral exhibits explicit dependence upon the electronic density of states (DOS) $g(\ep)$.  Limitations of the electron-phonon and electron-electron integrals are discussed in light of prior experimental and/or theoretical studies.  The simplicity of the integrals make them readily applicable to modeling experimental data.

(ii)  We thence apply our BTE expressions to gain new insight into three distinct experimental studies of Au:  femtosecond thermionic emission \cite{wang1994}, transient two-color photoemission \cite{Fann1992}, and subpicosecond Raman scattering \cite{Huang2014}.  We emphasize that none of these experiments has previously been analyzed using the BTE.  Our modeling of the fs thermionic-emission experiment shows that the high energy tail of the distribution -- which is responsible for the emission -- can be significantly hotter than that predicted by the 2T model.  Our calculations appropriate to both the photoemission and Raman scattering studies reveal that these experiments are not independently sensitive to both electron-phonon scattering and laser-excitation level.  However, they are both quite sensitive to electron-electron scattering.  These observations are largely the result of the timescales of both experiments being (predominantly) in the subpicosecond range.  Using a well established value for the electron-phonon scattering strength [see discussion of Table \ref{table3b}], our analysis of both experiments suggests $\tau_{ee}$(1$\,$eV) has a value in the range of 25 to 55 fs.  This result is in line with previous analyses of ballistic-electron emission (BEE) data \cite{Bauer1993,Bell1996,Reuter1999,Reuter2000,dePablos2003}, but such values are somewhat smaller than those extracted from time resolved two-photon photoemission (TR2PPE) experiments \cite{Aeschlimann1996,Cao1998,Aeschlimann2000,Bauer2015}.

\bigskip

\appendix

\section{Relationship of $\langle \Omega^n \rangle$ to moment Debye temperatures $\Theta_m$}
\label{Appendix B}

As is evident in our discussion of electron-phonon scattering, a key quantities is the moment $\langle \Omega^2 \rangle$ associated with the spectral function $\alpha^2 F(\Omega)$.  In the literature $\langle \Omega^2 \rangle$ has been estimated using 
\be{\eq}
\label{B1}
\hbar^2 \langle \Omega^2 \rangle = \frac{1}{2} k_B^2 \Theta_D^2,
\en{\eq}
where $\Theta_D$ is a Debye temperature \cite{Allen1972,Papa1977,Papa1981,Allen1987b,Sanborn1989,Medvedev2020}.  This expression, however, begs a couple of questions.  First, for a given metal, which Debye temperature should we use?  Second, under what approximations is Eq.~(\ref{B1}) valid?  We now answer each of these questions and along the way derive a general relationship between $\langle \Omega^n \rangle$ and the moment Debye temperatures $\Theta_m$ of a solid.

Experimental Debye temperatures can be extracted from a variety of measurable quantities, including specific heat, entropy, elastic constants, and atomic mean-squared displacements.  If a solid were to somehow have a pure Debye DOS (i.e., one strictly proportional to $\Omega^2$), then all of these different measurements would -- in principle -- yield the same value for $\Theta_D$.  However, because no solid exhibits such a simple spectrum, measured properties do not all yield the same Debye temperatures.

\begin{figure}[b]
\centerline{\includegraphics[scale=0.60]{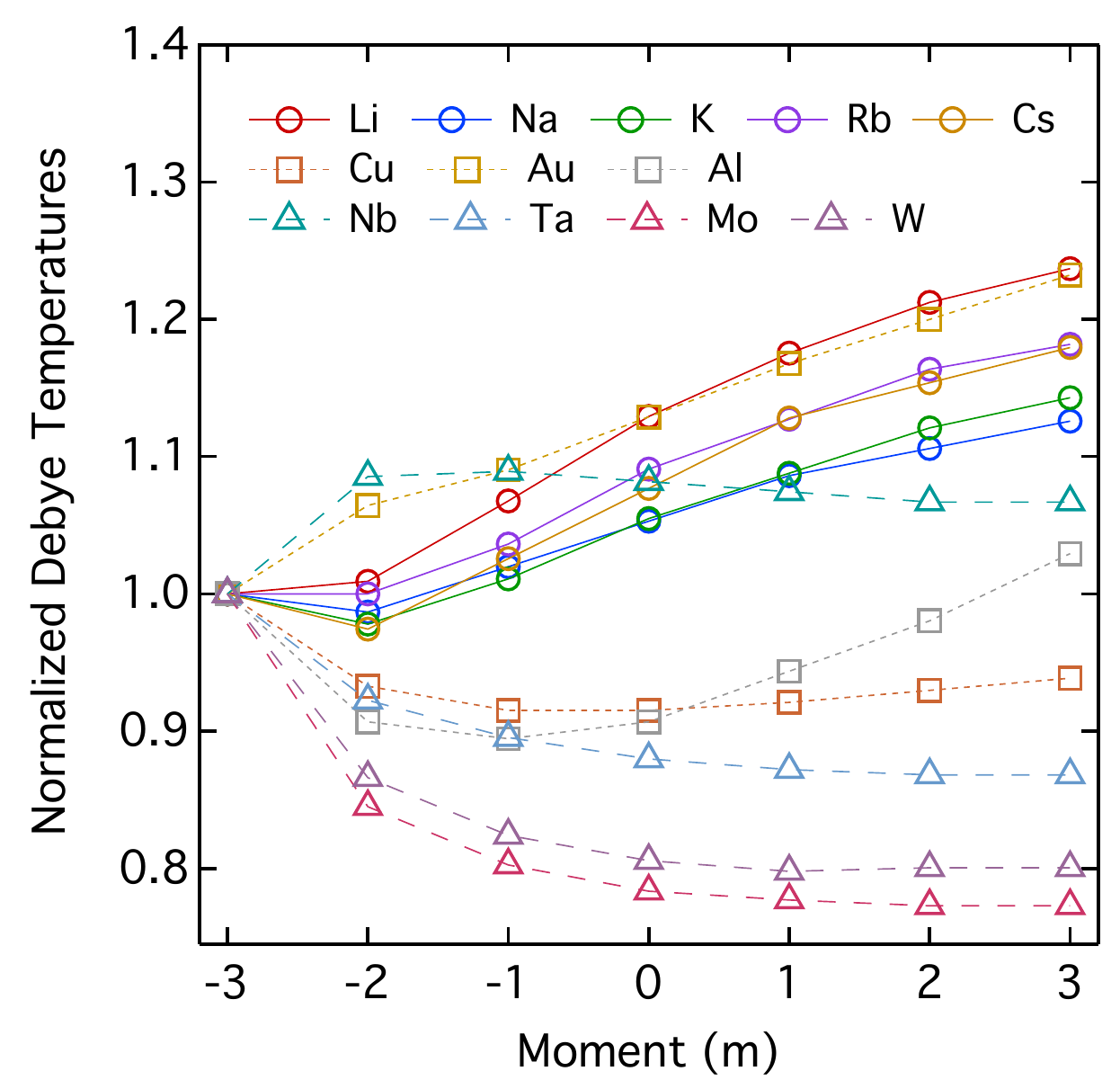}}
\caption{Moment Debye temperature ratios $\Theta_m / \Theta_{-3}$ for $m = -3$ to 3 for the metals indicated.  Values of $\Theta_m$ are sourced from Ref.~[\onlinecite{Wilson2011}].}
\label{FigB1}
\end{figure}

As it turns out, each of the experimental quantities mentioned above is related to moment Debye temperatures $\Theta_m$ associated with the vibrational DOS $F(\Omega)$ of the solid.  For $m = -2, -1, 1, 2, 3, ...$ these Debye temperatures are directly given by \cite{Grimvall1981}
\be{\eq}
\label{B2}
\Theta_m = \bigg( \! \frac{\hbar}{k_B} \! \bigg) \, \bigg[ \frac{m+3}{3} \frac{\int d\Omega \, F(\Omega) \, \Omega^m}{\int d\Omega \, F(\Omega) }\bigg]^{1/m}.
\en{\eq}
For $m = 0$ this expression is indeterminate.  The limit $m \rightarrow 0$ is, however, well defined and given by 
\be{\eq}
\label{B3}
\Theta_0 = \frac{\hbar \omega_0}{k_B} \exp\bigg( \frac{1}{3} +  \frac{\int \ln(\omega/\omega_0) g(\omega) d\omega}{\int g(\omega) d\omega} \bigg),
\en{\eq} 
where $\omega_0$ is an arbitrary frequency.  Because in general $F(\Omega) \sim \Omega^2$ at the lowest frequencies, Eq.~(\ref{B1}) is also indeterminate for $m = -3$.  However, it can be shown the $m \rightarrow -3$ limit is also well defined.  

So which $\Theta_m$'s are related to which measurements?  The low and high temperature limits of the specific heat are governed by the $m = -3$ and $m = 2$ moments, respectively.  To be clear, this means that when the vibrational contribution to the specific heat at low (high) temperatures is interpreted in terms of the Debye model of vibrations, it is $\Theta_{-3}$ ($\Theta_2$) that is extracted from the data.  Similarly, the low (high) temperature limit of the entropy is governed by the $m = -3$ ($m = 0$) moment, while the low (high) temperature limit of the mean-squared amplitude of vibration is governed by the $m = -1$ ($m = -2$) moment.  Elastic constants, owing to their direct relationship to the very lowest vibrational frequencies, are also directly related to $\Theta_{-3}$.  

We now relate the moments $\langle \Omega^n \rangle$ to these Debye temperatures.  We first note  Eq.~(\ref{9}) implies
\be{\eq}
\label{B4}
\langle \Omega^n \rangle = \frac{\lambda \langle \Omega^n \rangle}{\lambda} = \frac{\int d\Omega \, \alpha^2 F(\Omega) \, \Omega^{n-1}}{\int d\Omega \, \alpha^2 F(\Omega) \, \Omega^{-1}}.
\en{\eq}
For many metals it has been observed the $\Omega$ dependence of $\alpha^2 F(\Omega)$ is largely contained within $F(\Omega)$ \cite{dynes1968,Rowell1969,Truant1972,Beaulac1982,Savrasov1996,Liu1999}.  To good approximation we may thus write
\be{\eq}
\label{B5}
\langle \Omega^n \rangle  = \frac{\int d\Omega \, F(\Omega) \, \Omega^{n-1}}{\int d\Omega \, F(\Omega) \, \Omega^{-1}}.
\en{\eq}
Using Eqs.~(\ref{B2}) and (\ref{B5}) it is straightforward to show for $n = -1, 0, 2, 3, ...$
\be{\eq}
\label{B6}
\hbar^n \langle \Omega^n \rangle  =  \frac{2}{n+2} \, k_B^n \, \Theta_{n-1}^{n-1} \, \Theta_{-1},
\en{\eq}
and for $n = 1\,$ \footnote{While Eq.~(\ref{B7}) can be simply obtained from Eq.~(\ref{B6}) by setting $n = 1$, it is best to express this case separately, owing to $\Theta_0$ not being directly given by Eq.~(\ref{B2}).}
\be{\eq}
\label{B7}
\hbar \langle \Omega \rangle  =  \frac{2}{3}  \, k_B  \, \Theta_{-1}.
\en{\eq}
We note for $n = 2$ Eq.~(\ref{B6}) yields
\be{\eq}
\label{B8}
\hbar^2 \langle \Omega^2 \rangle  =  \frac{1}{2}  \, k_B^2 \, \Theta_{1}  \Theta_{-1}.
\en{\eq}
By comparing Eq.~(\ref{B8}) with Eq.~(\ref{B1}) we see that if moment Debye temperatures are available, then $\Theta_D^2$ in Eq.~(\ref{B1}) should be replaced by $\Theta_{1}  \Theta_{-1}$.

These results beg one final question.  How important is it to use the appropriate Debye temperatures when estimating $\langle \Omega^n \rangle$?  We answer this question by comparing moment Debye temperatures for a selection of elemental metals.  In Table \ref{table2} we list the moment Debye temperatures for 12 metals, while in Fig.~\ref{FigB1} we plot the ratios $\Theta_m / \Theta_{-3}$ for $m = -3$ to 3.   Given the results as displayed in the figure, we surmise it is generally not a good idea to use the commonly quoted $\Theta_{-3}$ (typically obtained from low-temperature specific-heat measurements) as a substitute for any of the higher-order Debye temperatures.  For example, for Li and Au $\Theta_1$ is $\sim \,$17 \% higher than $\Theta_{-3}$, while for W $\Theta_1$ is $\sim \,$20 \% lower than $\Theta_{-3}$.  On the other hand, for some metals $\Theta_m$ does not vary strongly with $m$ over a range of $m$ values.  For example, for the transition metals $\Theta_m$ does not substantially vary for $n \ge -2$.

Relevant to our analyses of experimental Au data in Sec.~\ref{SectionV}, we find  $\hbar^2 \langle \Omega^2 \rangle =$ 113.6 meV$^2$, which is obtained using Eq.~(\ref{B8}) and the values $\Theta_{-1} =$ 169 K and $\Theta_{1} =$ 181 K from Table \ref{table2}.

\begin{table*}[t]
\caption{Moment Debye temperatures for a selection of elemental metals.  Also shown is the minimum lattice temperature $T_p^{(min)}$ for which the linear $\hbar \Omega$ approximation to $(d \langle \ep \rangle / dt)_{ep}$ is valid. \label{table2}}
\vspace{0.2cm}
\begin{tabular}{l@{\hspace{0.5cm}}r@{\hspace{0.5cm}}r@{\hspace{0.5cm}}r@{\hspace{0.5cm}}r@{\hspace{0.5cm}}r@{\hspace{0.5cm}}r@{\hspace{0.5cm}}r@{\hspace{0.5cm}}r@{\hspace{0.5cm}}r}
\hline
\hline  
    
Metal &\multicolumn{7}{c} {Moment\footnote{$\Theta_i$'s are from Ref.~[\onlinecite{Wilson2011}]}} & $T_p^{(min)}$ \\

& $-3$  & $-2$ & $-1$ & $0$ & $1$ & $2$ & $3$ & (K)  \\

\hline

Li 			& 325 	& 328	& 347	& 367	& 382 	& 394	& 402	& 307	  \\	

Na  		 	& 151	&149 	& 154	& 159	& 164 	& 167	& 170	& 129		  \\

K  		  	& 91		& 89		& 92		& 96		& 99	 	& 102	& 104	& 79		 \\

Rb 		  	& 55		& 55		& 57		& 60		& 62	 	& 64		& 65	 	& 50			 \\

Cs 		  	& 39		& 38		& 40		& 42		& 44	 	& 45		& 46	 	& 35			 \\

Cu 		  	& 342	& 319	& 313	& 313	& 315 	& 318	& 321 	& 242		 \\

Au 		  	& 155	& 165	& 169	& 175	& 181 	& 186	& 191 	& 146			 \\

Al 		  	& 408	& 370	& 365 	& 370	& 385 	& 400	& 420 	& 327		 \\

Nb 		  	&269		& 292	& 293	& 291	& 289 	& 287	& 287 	& 213		 \\

Ta 		  	& 258	& 238	& 231	& 227	& 225 	& 224	& 224 	& 167		 \\

Mo 		  	& 471	& 398	& 378	& 369	& 366 	& 364	& 364 	& 271		 \\

W 		  	& 381	& 330	& 314	& 307	& 304 	& 305	& 305	& 228		 \\

\hline
\hline

\end{tabular}
\smallskip
\end{table*}

\section{Higher-order corrections to the electron-phonon interaction}
\label{Appendix A}

In Sec.~\ref{Sec11.A} we derive expressions for $(df/dt)_{ep}$ and $(d \langle \ep \rangle / dt)_{ep}$ to first order in $\hbar \Omega$; respectively these expressions are given by Eqs.~(\ref{8}) and (\ref{11}).  Here we present the next nonzero term in the $\hbar \Omega$ expansion of $(df/dt)_{ep}$.  Then, by considering $(d \langle \ep \rangle / dt)_{ep}$ for constant $g(\ep)$, we discuss the conditions under which the expansion to linear order in $\hbar \Omega$ is sufficient.

Because all even-powered terms in the expansion of $(df/dt)_{ep}$ are zero, the next nonzero term is the $(\hbar \Omega)^3$ term, which can be written as
\begin{align}
\label{A1}
\bigg(& \! \frac{\partial f(\ep)}{\partial t} \! \bigg)_{\!\! ep}^{\!\! (\Omega^3)} = \frac{\pi \hbar ^3\, \lambda \langle \Omega^4 \rangle}{g(\ep_F)} \nonumber \\
	&\times \bigg[ \frac{g(\ep)}{12} \bigg( 2 f''' - 4 f f''' + \frac{f''}{k_B T_p} + f'''' k_B T_p \bigg) \nonumber \\
	&+ \frac{g'(\ep)}{6} \bigg( 3 f'' - 6 f f'' + \frac{f'}{k_B T_p} + 2 f''' k_B T_p \bigg) \nonumber \\
	&+ \frac{g''(\ep)}{2} \big( f' - 2 f f' + f'' k_B T_p  \big) \nonumber \\
	&+ \frac{g'''(\ep)}{3} \big( f - f^2 + f' k_B T_p  \big) \bigg],
\end{align}
where for compactness we have suppressed the argument $\ep$ of $f$ and its derivatives.  This equation could certainly be inserted into the integral on the right side of Eq.~(\ref{11}) and thence used to numerically calculate $(d \langle \ep \rangle / dt)_{ep}^{(\Omega^3)}$, but our main goal here is to get a sense of when we are justified in solely utilizing the $\hbar \Omega$ term.

To this end we now calculate $(d \langle \ep \rangle / dt)_{ep}^{(\Omega^3)}$ under the simplifying assumption of constant $g(\ep) = g(\ep_F)$.  In this case Eq.~(\ref{A1}) readily simplifies to
\begin{align}
\label{A2}
\bigg( \! \frac{\partial f(\ep)}{\partial t} \! &\bigg)_{\!\! ep}^{\!\! (\Omega^3)} = \frac{\pi \hbar ^3\, \lambda \langle \Omega^4 \rangle}{12} \nonumber \\
	&\times \bigg( 2 f''' - 4 f f''' + \frac{f''}{k_B T_p} + f'''' k_B T_p \bigg).
\end{align}
If we now (i) substitute this relation into Eq.~(\ref{10}) and (ii) liberally utilize integration by parts we end up with the tidy expression
\begin{align}
\label{A3}
\bigg( \! \frac{\partial \langle \ep \rangle}{\partial t} \! &\bigg)_{\!\! ep}^{\!\! (\Omega^3)} = \frac{\pi \hbar^3 \, \lambda \langle \Omega^4 \rangle g(\ep_F)}{12} \nonumber \\
	&\times \bigg[ \frac{1}{k_B T_p  } - 6 \int d \ep \, (f')^2 \bigg].
\end{align}
We note under the same conditions Eq.~(\ref{11}) simplifies to
\begin{align}
\label{A4}
\bigg( \! \frac{\partial \langle \ep \rangle}{\partial t} \! &\bigg)_{\!\! ep}^{\!\! (\Omega)} = \pi \hbar\, \lambda \langle \Omega^2 \rangle g(\ep_F) \nonumber \\
	&\times \bigg[ k_B T_p  -  \int d \ep \, f (1 - f) \bigg].
\end{align}
A cursory comparison of these last two equations indicates the ratio of the magnitudes of these two terms might be estimated using
\begin{equation}
\label{A5}
\bigg| \frac{(d \langle \ep \rangle / dt)_{ep}^{(\Omega^3)}}{(d \langle \ep \rangle / dt)_{ep}^{(\Omega)}} \bigg| \approx \frac{\hbar^2 \langle \Omega^4 \rangle}{12 \langle \Omega^2 \rangle k_B^2 T_p^2},
\end{equation}
which is consistent with the statement in Sec.~\ref{Sec11.A} that only the first-order term is necessary if $k_B T_p \gtrsim \hbar \Omega$.

For a more quantitative comparison of Eq.~(\ref{A3}) and (\ref{A4}) we now assume $f(\ep)$ is a thermal distribution described by a temperature $T_e$.  The results of Sec.~\ref{Sec11.A} show that Eq.~(\ref{A4}) then simplifies to
\begin{equation}
\label{A6}
\bigg( \! \frac{\partial \langle \ep \rangle}{\partial t} \! \bigg)_{\!\! ep}^{\!\! (\Omega)} = \pi \hbar\, \lambda \langle \Omega^2 \rangle g(\ep_F) k_B \, (T_p - T_e).
\end{equation}
Similarly, using $\int d\ep \, (f'_{\scriptscriptstyle \! F\!D})^2 = 1 / (6 k_B T_e)$, Eq.~(\ref{A3}) becomes
\begin{equation}
\label{A7}
\bigg( \! \frac{\partial \langle \ep \rangle}{\partial t} \! \bigg)_{\!\! ep}^{\!\! (\Omega^3)} = \frac{\pi \hbar^3 \, \lambda \langle \Omega^4 \rangle g(\ep_F)}{12 k_B} \bigg( \frac{1}{T_p} - \frac{1}{T_e} \bigg).
\end{equation}
Using these last two relations we find
\begin{equation}
\label{A8}
\frac{(d \langle \ep \rangle / dt)_{ep}^{(\Omega^3)}}{(d \langle \ep \rangle / dt)_{ep}^{(\Omega)}}  = - \frac{\hbar^2 \langle \Omega^4 \rangle}{12 \langle \Omega^2 \rangle k_B^2 T_p T_e}.
\end{equation}
Insofar as a laser excited metal has $T_e \ge T_p$, this relationship is somewhat less restrictive than that suggested by Eq.~(\ref{A5}).  We note Eq.~(\ref{A8}) is implicit in Eq.~(14) of Ref.~[\onlinecite{allen1987}].

We now apply Eq.~(\ref{A8}) to the set of metals listed in Table \ref{table2}.  Using the results of Appendix \ref{Appendix B} that relate $\langle \Omega^n \rangle$ to $\Theta_m$, we can rewrite Eq.~(\ref{A8}) as
\begin{equation}
\label{A9}
\frac{(d \langle \ep \rangle / dt)_{ep}^{(\Omega^3)}}{(d \langle \ep \rangle / dt)_{ep}^{(\Omega)}}  = - \frac{\Theta_3^3}{18 \, \Theta_1  T_p^2}
\end{equation}
under conditions of low electronic excitation ($T_e \approx T_p$).  It should not be unreasonable to simply use the linear $\hbar \Omega$ aproximation in calculating the electron-phonon dynamics as long as the magnitude of Eq.~(\ref{A9}) is less than $\sim$ 0.1.  We thus calculate the value of $T_p$ that results in Eq.~(\ref{A9}) equaling $-$0.1 for each metal; we designate this temperature $T_p^{(min)}$.  The results, presented in the last column of Table \ref{table2}, show that for most metals a room temperature value of $T_p$ is entirely adequate.  Possible exceptions from this list of metals are Li and Al.


\section*{References}


\vspace{-0.5cm}

\bibliography{AlkaliMetals}

\end{document}


\title{Supplemental Material for ``Excitation and Relaxation of Nonthermal Electron Energy Distributions in Metals with Application to Gold''}

\author{D. M. Riffe}

\affiliation{Physics Department, Utah State University, Logan, UT 84322, USA}

\author{Richard B. Wilson}

\affiliation{Mechanical Engineering Department and Materials Science and Engineering Department, University of California, Riverside, CA 92521, USA}


\maketitle

Figure \ref{Fig1} plots electron distribution functions obtained from (i) all of the reported transient photoemission measurements of Fann \textit{et al.} \cite{Fann1992} and (ii) our model calculations, as discussed in Sec.~IV.B of the paper.  As described there in conjunction with Fig.~3, parameters in the model calculations were adjusted to provide the best match to the early-time data (0 - 400 fs for the high-fluence data and 0 - 130 fs for the low-fluence data) in the energy range $0 < \delta \epsilon < 0.8$ eV.  These best-match fits are plotted in Fig~\ref{Fig1}.  As discussed in the paper and shown Fig.~\ref{Fig1}, the early-time data are well described by the model, but the data at later times generally fall below the calculations.  We find that no set of parameters are able to adequately describe both the early-time and late-time data.

\bigskip

\renewcommand{\thefigure}{S\arabic{figure}}

\begin{figure*}[h]
\centerline{\includegraphics[scale=0.385]{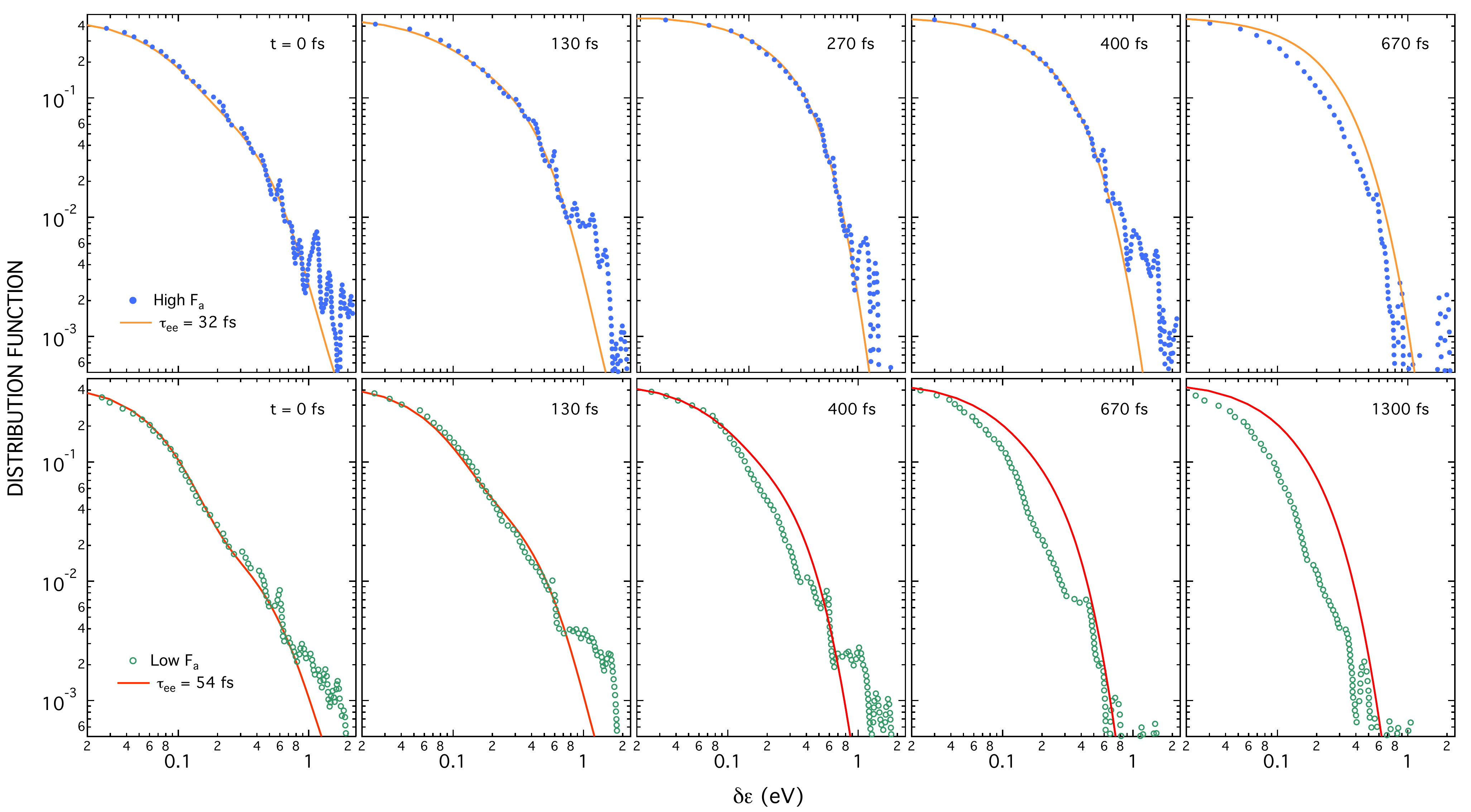}}
\caption{Electron distribution functions from (i) transient photoemission data of Fann \textit{et al.}\cite{Fann1992} (circles) and (ii) our model calculations (solid lines).  The parameter $\tau_{ee}$ is the single-electron scattering time at $\delta \epsilon =$ 1 eV.  High $F_a$ and low $F_a$ refer to experimental absorbed fluence of 300 $\pm$ 90 and 120 $\pm$ 40 $\mu$J, respectively.  See text for further details.}
\label{Fig1}
\end{figure*}

\bigskip


\bibliography{AlkaliMetals}